\documentclass[aps,prb,twocolumn,reprint,superscriptaddress,longbibliography]{revtex4-2}

\usepackage{ae,aecompl}
\usepackage[T1]{fontenc}
\usepackage[utf8]{inputenc}
\usepackage{babel}
\usepackage{float}
\usepackage{amsmath}
\usepackage{amssymb}
\usepackage{graphicx}
\usepackage{wasysym}
\usepackage{amsmath}
\usepackage{soul}
\setstcolor{red}
\usepackage[export]{adjustbox}
\usepackage{color}
\usepackage[normalem]{ulem}
\usepackage[unicode=true,bookmarks=true,bookmarksnumbered=false,bookmarksopen=false,breaklinks=false,pdfborder={0 0 1},backref=false,colorlinks=true,linkcolor=magenta,citecolor=blue]{hyperref}

\begin{document}

\global\long\def\ket#1{\left|#1\right\rangle }%
\global\long\def\bra#1{\left\langle #1\right|}%
\global\long\def\braket#1#2{\langle#1|#2\rangle}%
\global\long\def\expectation#1#2#3{\langle#1|#2|#3\rangle}%
\global\long\def\average#1{\langle#1\rangle}%

\title{Thermal avalanches in isolated many-body localized systems}

\author{Muhammad Sajid}
\thanks{The first two authors have contributed equally to this paper.}
\affiliation{Institute of Fundamental and Frontier Sciences, University of Electronic Science and Technology of China, Chengdu 611731, P. R. China}
\affiliation{Key Laboratory of Quantum Physics and Photonic Quantum Information, Ministry of Education, University of Electronic Science
and Technology of China, Chengdu 611731, P. R. China}

\author{Rozhin Yousefjani}
 \email{ryousefjani@hbku.edu.qa}
\affiliation{Qatar Center for Quantum Computing, College of Science and Engineering, Hamad Bin Khalifa University, Doha, Qatar}

\author{Abolfazl Bayat}
\email{abolfazl.bayat@uestc.edu.cn}
\affiliation{Institute of Fundamental and Frontier Sciences, University of Electronic Science and Technology of China, Chengdu 611731, P. R. China}
\affiliation{Key Laboratory of Quantum Physics and Photonic Quantum Information, Ministry of Education, University of Electronic Science
and Technology of China, Chengdu 611731, P. R. China}
\affiliation{Shimmer Center, Tianfu Jiangxi Laboratory, Chengdu 641419, China}
\date{\today}
%
\begin{abstract}
Many-body localization is a profound phase of matter affecting the entire spectrum which emerges in the presence of disorder in interacting many-body systems. Recently, the stability of many-body localization has been challenged by the avalanche mechanism, in which a small thermal region can spread, destabilizing localization and leading to global thermalization of the system. A key unresolved question is the critical competition between the thermal region's influence and the disorder strength required to trigger such an avalanche. Here, we numerically investigate many-body localization stability in an isolated Heisenberg spin chain of size $L$ subjected to a disordered magnetic field. By embedding a tunable thermal region of size $P$, we analyze the system's behavior in both static and dynamical regimes using entanglement entropy and the gap ratio. Our study yields two main findings. Firstly, for strong disorder, the avalanche only occurs if the thermal region scales with system size, specifically when $P/L$ exceeds a threshold value. Secondly, at strong disorder, we identify an intermediate phase between many-body localization and ergodic behavior as $P$ increases. This intermediate phase leaves its fingerprint in both static and dynamic properties of the system and tends to vanish in the thermodynamic limit. 
Although our simulations are restricted to finite system sizes, the analysis suggests that these results hold in the thermodynamic limit for isolated many-body systems.
\end{abstract}

\maketitle
\section{Introduction} \label{Intro}
The presence of disorder in interacting many-body systems results in the emergence of an interesting phase of matter known as many-body localization (MBL) which affects the entire spectrum of the system~\cite{Rev1,Rev2,Rev3,Rev4,Rev5,Rev6,imbrie2016many}. 
Over the past decade, many distinctive features of MBL have been identified, including the breakdown of eigenstate thermalization hypothesis~\cite{srednicki1994chaos,rigol2008thermalization,panda2020can}, the emergence of quasi-local integrals of motion~\cite{serbyn2013local,huse2014phenomenology}, Poisson-like level statistics~\cite{Huse1,Huse2,Huse3,Laflorencie1,yousefjani2023floquet}, the emergence of mobility edge~\cite{Laflorencie1,luschen2018single,yousefjani2023mobility}, area-law entangled eigenstates~\cite{Area-law1,Area-law2}, and logarithmically slow entanglement growth~\cite{Laflorencie2,UnboundedEG,EGSR1,EGSR2,EGSR3,EGSR4,EGSR5}. Thanks to recent advances in quantum technologies, various aspects of MBL have also been tested experimentally on various physical platforms~\cite{coldatoms1,coldatoms2,coldatoms3,coldatoms4,
2DRDAtoms,CriticalRbatom,Rbatom,Iontrap1,Iontrap2,Iontrap3,superconducting1,superconducting2,
superconducting3,superconducting4,
superconducting5,superconducting6,
superconducting7,superconducting8,
superconducting9,nitrogenvacancy1,nitrogenvacancy2,photonic,guo2023observation,shtanko2025uncovering}. Nonetheless, the stability of the MBL phase in the presence of a small thermal region in the system has been challenged by the avalanche theory~\cite{thiery2018many,de2017stability}. Thus, a key open question is to identify the criteria, likely dependent on both the size of the thermal region and the disorder strength, that trigger an avalanche, thereby destabilizing the MBL phase.

Analytical studies and renormalization group approaches indicate that quantum avalanches can arise when small ergodic regions or thermal inclusions act as local baths, gradually absorbing the surrounding localized degrees of freedom~\cite{thiery2018many,dumitrescu2019kosterlitz,gopalakrishnan2019instability,rubio2019many,SierantPRB2024,thiery2018many,de2017stability,de2017stability,LuitzAvalanch,colbois2024interaction}. 
In this scenario, MBL is not a genuine thermodynamic phase but only a prethermal regime that breaks down at sufficiently long times~\cite{sels2021dynamical,wu2024atomic,Suntajs2020,ColmenarezAvlnch2024} or large system sizes~\cite{PeacockPRB}. 
The avalanche mechanism has been investigated via two primary frameworks:
(i) isolated systems, where weak (or zero) disorder defines the thermal region~\cite{goihl2019exploration,PeacockPRB,SierantPRB2024,brighi2023many}; and (ii) open quantum systems, where the thermal region couples to an external bath~\cite{ScoccoPRB2024,MorningstarPRB2022,PhysRevB.106.L020202,tu2023avalanche}.  
Experimental evidence for avalanche-like behavior has recently been observed in cold-atom systems~\cite{Lonard2023}.

\begin{figure}[t!]
    \includegraphics[width=0.48\textwidth]{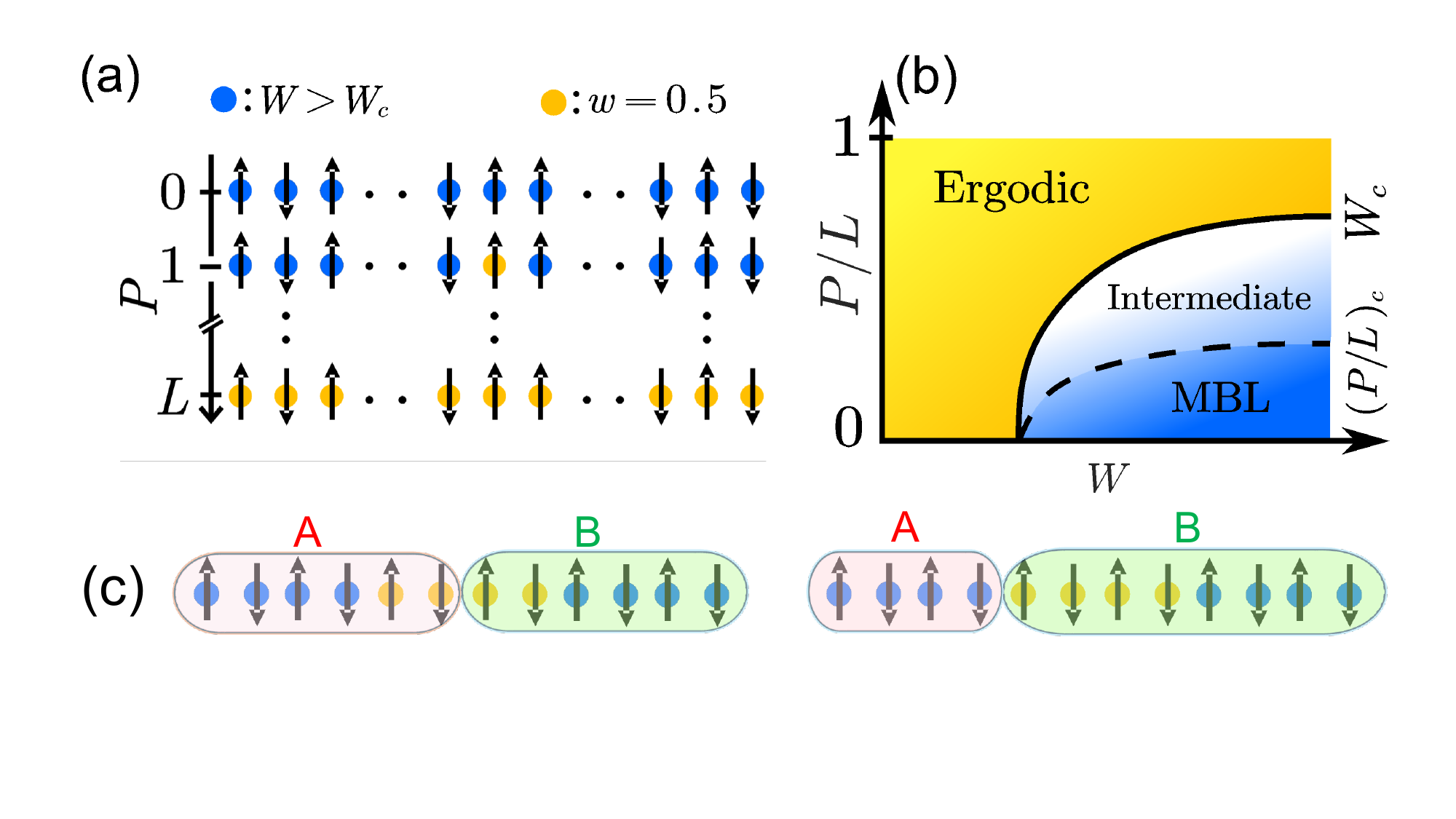}
    \caption{ (a) Schematic of our model system. While the dark points are exposed to strong random disorder, the light points are subjected to a weak random disorder of strength $w{=}0.5$.
    (b) Graphical visualization of the phase diagram as a function of disorder strength $W$ and the size of the weak-disorder region $P{/}L$. The solid (dashed) line indicates the phase boundary controlled by the critical $W$ ($P{/}L$). The intermediate area gives a rough approximation of the region wherein the MBL is unstable in the thermodynamic limit.
     (c) Schematic of two different types of chain partitioning for calculation of the entanglement entropy.}
    \label{fig:1}
\end{figure}

In contrast, a parallel line of research argues that MBL can persist even in systems with avalanche-inducing features, provided that certain conditions are satisfied~\cite{colbois2024interaction,goihl2019exploration,MorningstarPRB2022,ScoccoPRB2024,foo2023stabilization,decker2022many}.
These studies show that localization instability is not universal and may depend sensitively on system architecture~\cite{rubio2019many,potirniche2019exploration,foo2023stabilization,decker2022many,doggen2020slow}, interaction range~\cite{dobrosavljevic2005absence,rademaker2020slow}, or even the characteristics of the thermal regions~\cite{vojta2006rare}. For instance, while a fixed-size bath that is described by a suitable random matrix predicts the instability of the MBL~\cite{LuitzAvalanch}, the presence of the same size thermal region in the Heisenberg chain may not destabilize the MBL phase~\cite{goihl2019exploration}.
Indeed, to resolve these conflicts a systematic analysis of MBL instability under the presence of a thermal region is highly desirable.

Here, we address this issue by numerically investigating the stability of the MBL in an isolated Heisenberg spin chain of size $L$ subjected to a random magnetic field. 
We intentionally embed a thermal region of $P$ spins with weak-disorder and analyze the system’s static and dynamical behavior using entanglement entropy and the gap ratio.
Our results show that avalanches occur in strongly disordered systems only when the thermal region scales with system size and particularly exceeds a threshold value of $P{/}L{>}0.1$.
As $P$ increases, an intermediate phase emerges between the MBL and ergodic regimes, manifesting in both static and dynamic properties of the system. However, this phase is expected to vanish in the thermodynamic limit. Fig.~\ref{fig:1}(b)  presents a graphical visualization of the extracted phase diagram as a function of disorder strength $W$ and the relative size of the thermal region $P{/}L$. The solid (dashed) line approximates the phase boundaries controlled by disorder strength $W$ (relative size of the thermal region $P{/}L$).

\section{Model and methodology} \label{model}
We consider a Heisenberg spin-$1/2$ chain of size $L$ in the presence of disordered magnetic fields as  
\begin{align}\label{eq:_1}
H = J\sum_{i = 1}^{L-1} \boldsymbol{\sigma}_i \cdot \boldsymbol{\sigma}_{i+1}  + W\sum_{i\notin\mathcal{P}} r_i \sigma_i^{z} + w\sum_{i\in\mathcal{P}} r_i \sigma_i^{z},
\end{align}
where, $J{=}1$ is the exchange coupling, $\boldsymbol{\sigma}_i{=}(\sigma_i^x, \sigma_i^y, \sigma_i^z)$ is the vector of Pauli matrices, $r_i$'s are random numbers drawn from a uniform distribution $[-1,1]$, $w{=}0.5$ is a fixed weak disorder strength which affects the $P$ spins at the center of the chain (which we call it region $\mathcal{P}$) and $W$ is a controllable disorder strength which affects the spins out of the region $\mathcal{P}$. See Fig.~\ref{fig:1}(a) for the schematic of the system. 
Since the Hamiltonian conserves total magnetization $\sigma^{z}_{\mathrm{tot}}{ = }\sum_i \sigma_i^z$, we restrict our numerical analysis to the special sector of the spectrum which is spanned by a set of eigenstates $\{|E_k\rangle\}$ such that $\bra{E_k}\sigma^{z}_{\text{tot}}\ket{E_k} {=} 0$.
We investigate the physics of the system by controlling two parameters, namely $W$ and $P$. 
Without the weak-disorder region $\mathcal{P}$, namely for the fully disordered chain, the system exhibits a phase transition at a specific disorder strength $W{=}W_c$ between ergodic and MBL phases. 
In the following, we investigate the MBL transition and its stability for $P{\geq}0$ through static and dynamical analysis.

\section{Static analysis} \label{static}
In the static case, we calculate two standard quantities, namely the gap ratio (GR) and bipartite entanglement entropy (EE), for 50 eigenenergies $E_k$ and the corresponding eigenstates ${\ket{E_k}}$ in the mid-spectrum $\varepsilon{=}0.5$ of the Hamiltonian $H$,
extracted using the exact diagonalization method after normalizing the energy spectrum as $\varepsilon{=}{(E_k-E_{\min})/(E_{\max}-E_{\min}})$ wherein $E_{\max}$ ($E_{\min}$) is the largest (smallest) eigenenergy of the system. 
For an ordered set of eigenenergies $\{E_k\}$, with energy gaps between consecutive levels $\delta_k{=}E_{k+1}{-}E_k$,
the ratio $r_k {=} \text{min}(\delta_{k+1},\delta_k)/\text{max}(\delta_{k+1},\delta_k)$ and its average $\overline{\langle r\rangle}$ over energy levels and sample realizations can
diagnose a global change in spectral statistics.
Strong repulsion between neighboring levels in the ergodic phase yields Gaussian distribution with $\overline{\langle r\rangle} {\simeq} 0.5307$.
However, the MBL phase exhibits Poisson distribution with $\overline{\langle r\rangle}{\simeq} 0.3863$.
As the EE behavior depends on the choice of partitioning, for each eigenstate $\ket{E_k}$, the EE can be calculated by partitioning the chain in different ways. In this work, we focus on the EE for two types of partitioning presented in Fig.~\ref{fig:1}(c).
In the first case, we cut the chain from the midpoint in a way that the subsystems $A$ and $B$ are equal in size. 
In the second scenario, we separate the strongly disordered region on the most left side from the rest of the chain, resulting in non-equal subsystems $A$ and $B$.  
In both cases, the EE is defined as
$S(\ket{E_k}) {=} {-}\text{Tr}[\rho_A^{k} \ln \rho_A^{k}]$,
where
$\rho_A^{k} {=} \text{Tr}_B[\ket{E_k}\bra{E_k}]$
is the reduced density matrix of subsystem $A$, obtained after tracing out subsystem $B$.
For clarity, we denote the EE in the first case with equal subsystems $A$ and $B$ as $S$ and for the second case with non-equal subsystems $A$ and $B$ as $S_{\mathrm{mbl}}$.   

In the ergodic phase, eigenstates are expected to resemble random pure states and exhibit volume-law EE consistent with Page entropy $S_{\mathrm{P}} {=} {\sum_{k= \mathcal{H}_B+1}^{\mathcal{H}_A\mathcal{H}_B} \ \frac{1}{k} {-} \frac{(\mathcal{H}_A -1 )}{2\mathcal{H}_B} }$, wherein $\mathcal{H}_A$ ($\mathcal{H}_B$) is the dimension of the Hilbert space associated with the subsystem $A$ ($B$)~\cite{PagePRL.71.1291,torres2016realistic}. 
Studies indicate that highly excited eigenstates in the MBL phase obey an area law for bipartite entanglement, due to the emergent integrability~\cite{Rev3,Area-law2}.
Therefore, averaged EE over eigenstates and disorder realizations $\overline{\langle S \rangle}$, results in $\overline{\langle S \rangle}/S_{\mathrm{P}} {\rightarrow} 1$ in the ergodic phase, and $\overline{\langle S \rangle}/S_{\mathrm{P}} {\rightarrow} 0$ in the MBL phase.

\begin{figure*}[t!]
\begin{centering}
\includegraphics[width=0.80\textwidth]{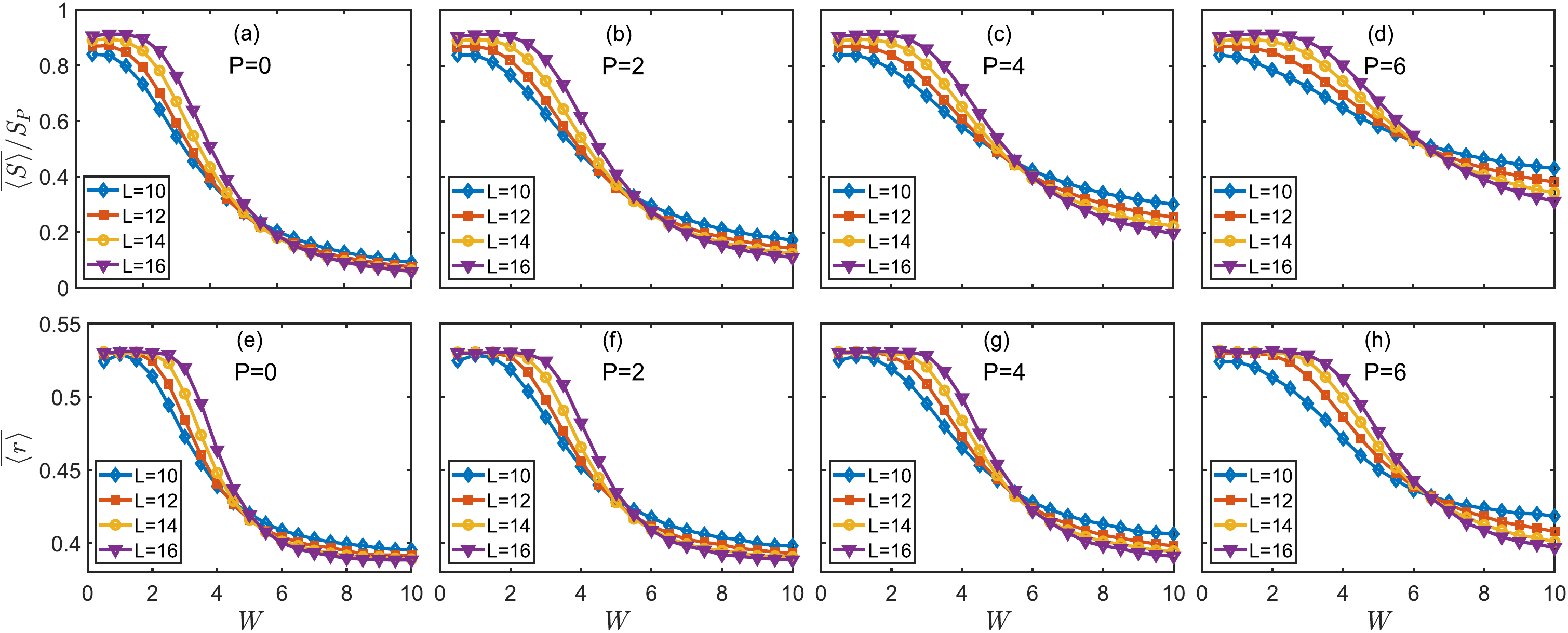} 
\end{centering}
\caption{First row: the averaged EE as a function of disorder strength $W$ in (a) fully disordered chains, and in (b-d) disordered chains with thermal regions of different sizes $P$ (indicated in the insets). The system sizes $L$ used in numerical simulation are indicated in the insets.  
Second row: the averaged GR as a function of disorder strength $W$ for the same set of $P$ values.
The results for EE and GR are obtained using $1000{-}2000$ and $10000{-}25000$ realizations, respectively.}
\label{fig:2}
\end{figure*}

The interplay between the MBL and ergodic phases is determined by two competing factors: (i) the disorder strength $W$; and (ii) the size of the thermal region $\mathcal{P}$, indicated by $P$ spins. While increasing $W$ tends to localize the system, increasing $P$ weakens the effect of disorder and tends to thermalize the system. In order to investigate these effects, we first fixed $P$ and vary $W$.
By increasing $W$, the system goes through a phase transition from the ergodic to the localized phase. Figures~\ref{fig:2}(a)-(d) depict the behavior of $\overline{\langle S \rangle}/S_{\mathrm{P}}$ as a function of the disorder strength $W$, for different system sizes $L{\in}[10,{\cdots},16]$, and various $P$'s.
As Fig.~\ref{fig:2}(a) shows, the system is ergodic for small disorders and $\overline{\langle S \rangle}/S_{\mathrm{P}}{\rightarrow}1$, indicating that the EE fulfills the volume-law entanglement, while increasing disorder $W$ leads to localization and $\overline{\langle S \rangle}/S_{\mathrm{P}}{\rightarrow}0$~\cite{dumitrescu2017scaling}.
The transition between these two phases happens at $W{=}W_c$. Thanks to scale invariant behavior at the phase transition point, one can utilize finite-size scaling analysis~\cite{pyfssa,Melchert2004,HoudayerPRB} to extract the critical value $W_c{\simeq}4.674 \pm 0.308$ as the prediction for the transition point in the thermodynamic limit, for a system with $P{=}0$. By embedding a thermal region, the critical disorder strength increases as can be noticed in Figs.~\ref{fig:2}(b)-(d) from the shift of the crossing points of the curves.
Similar analysis can be performed for the GR which is shown in Figs.~\ref{fig:2}(e)-(h). Decreasing GR from $\overline{\langle r\rangle} {\simeq} 0.5307$ to $\overline{\langle r\rangle} {\simeq} 0.3863$ clearly demonstrates the transition from the ergodic to the MBL phase. Here, finite-size scaling analysis results in $W_c{\simeq}4.871 \pm 0.243$ (for $P{=}0$), which is close to the value identified by the EE. 
Repeating the above analysis, using either EE or GR, we extract $W_c$ for various values of $P$ and report the results in the Appendix~(\ref{app.Pyfssa}).
A shift in $W_c$ with increasing $P$ is noticeable in the case of GR as well.
\begin{figure}[b!]
\begin{centering}
\includegraphics[width=0.48\textwidth]{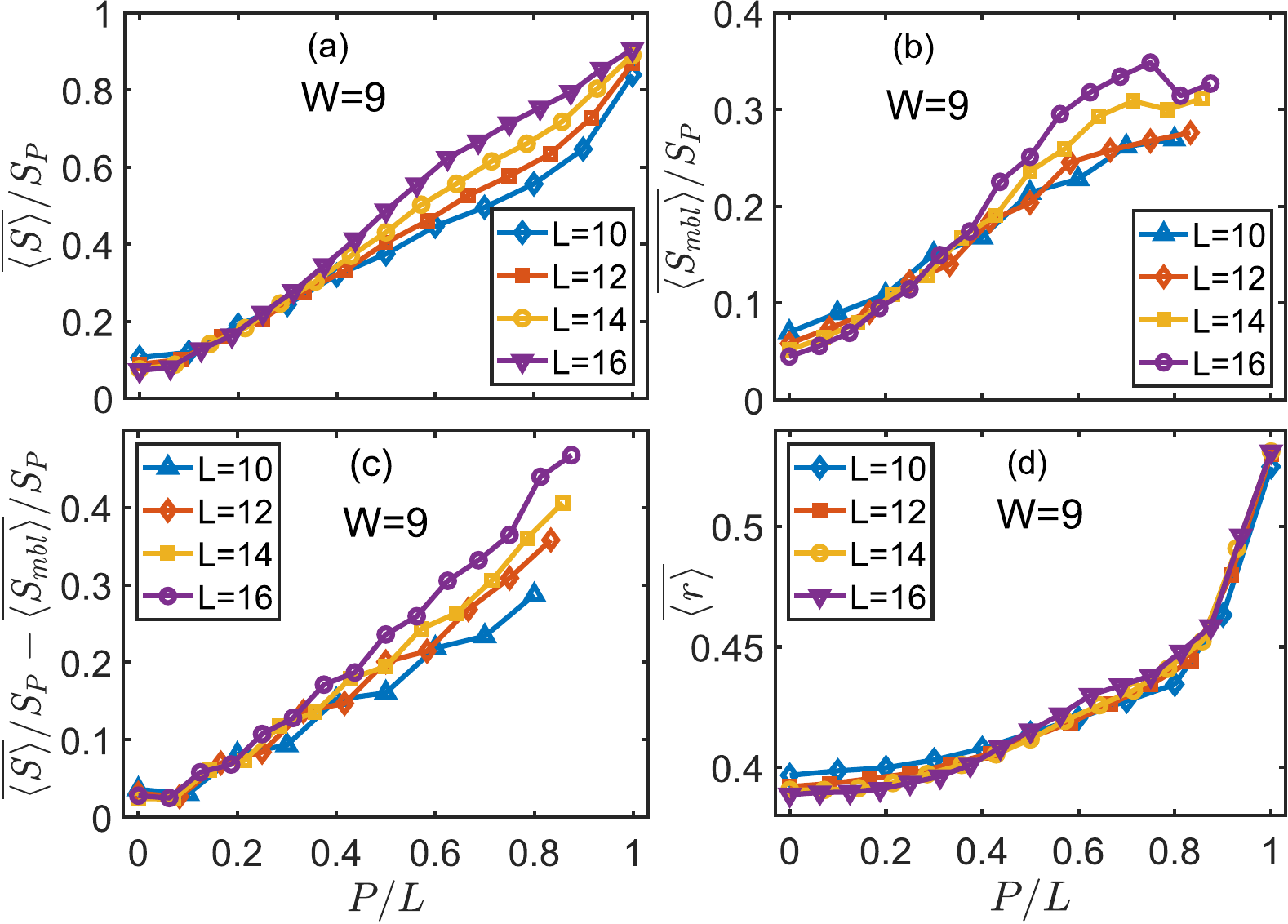} 
\end{centering}
\caption{(a) The averaged EE $\overline{\langle S \rangle}$, computed for half chain slicing as a function of $P{/}L$.  
(b) the averaged EE $\overline{\langle S_{\mathrm{mbl}} \rangle}$, computed by separating the strongly disordered region on the left side from the rest of the chain as a function of $P{/}L$. In both panels, the size of the system varies and $W{=}9$. (c) The difference between the results presented in (a) and (b). (d) The averaged GR as a function of $P{/}L$ for $W{=}9$.
The results for EE and GR are obtained using $1000{-}2000$ and $10000{-}25000$ realizations, respectively.}
\label{fig:2_second_part}
\end{figure}

Alternatively, one can fix disorder strength $W$ to a large value, enough for localizing the system for $P{=}0$, and then increase $P$ to thermalize the system gradually.  
In Fig.~\ref{fig:2_second_part}(a), we plot and $\overline{\langle S \rangle}/S_{\mathrm{P}}$ as a function of $P{/}L$, keeping $W{=}9$ fixed, for various system sizes. 
Several important features are observed.
First, regardless of the system size, MBL can survive the presence of a small weak-disorder region $\mathcal{P}$, evidenced by the early plateau of the curves. Second, after the initial plateau when $P{/}L{>}0.1$, the value of EE increases linearly by increasing $P/L$, such that the system eventually thermalizes to a volume law state.
Third, for $0.1{<}P{/}L{<}0.3$ the rate of change in the averaged EE diminishes by enlarging the system, ensuring the robustness of the MBL in the thermodynamic limit for the thermal region of size $P{>}1$.
Fourth, the rise of EE for $P{/}L{>}0.3$, is sharpened as the system size increases, suggesting that in the thermodynamic limit, thermalization occurs abruptly. In the calculation of $\overline{\langle S \rangle}$, we put the cut in the middle of the chain, which is inside the region $\mathcal{P}$, and one might be concerned that it captures only the EE in the low-disordered region, caused by the thermalization in this area.
To address this issue, we compute $\overline{\langle S_{\mathrm{mbl}} \rangle}$ 
by dividing the chain into non-equal subsystems, as illustrated on the right side of Fig.~\ref{fig:1}(c). This captures the EE behavior between the low-disordered area $\mathcal{P}$ and the strongly disordered part of the chain; see Fig.~\ref{fig:2_second_part}(b). Interestingly, $\overline{\langle S_{\mathrm{mbl}} \rangle}$ follows $\overline{\langle S \rangle}$ qualitatively. It shows the robustness of the MBL for $P{/}L{<}0.1$, slower growth of the EE for $0.1{<}P{/}L{<}0.3$, followed by sharper thermalization as the system size increases.
In Fig.~\ref{fig:2_second_part}(c), we plot $\overline{\langle S \rangle}{-}\overline{\langle S_{\mathrm{mbl}} \rangle}$, which shows the contributions to EE from interactions between $P$ spins, due to the thermalization.
As one can see, this portion of the EE shows the same behavior as $\overline{\langle S \rangle}$.
This analysis ensures that, although entanglement inside the thermal region is involved, $\overline{\langle S \rangle}$ can perfectly reveal the behavior of the system in terms of $P{/}L$ and $W$.

In Fig.~\ref{fig:2_second_part}(d), we depict $\overline{\langle r\rangle}$ versus $P{/}L$ in systems with different sizes, keeping the disorder strength fixed at $W{=}9$. The results provide further qualitative confirmation for the findings from the EE.     
Initially, the GR increases very slowly indicating the resilience of the MBL phase against small weak-disorder regions in all considered system sizes. The growth of the GR increases significantly as $P{/}L$ increases, again indicating the instability of the MBL for reasonably large weak-disorder regions. Similar to the EE case, the growth rate sharpens as the system size increases.  The above observations, for both GR and EE, lead us to the following conjecture: to trigger an avalanche and destabilize the MBL phase, the ratio $P/L$ should exceed a critical threshold, i.e.  $P/L{>}0.1$.

\begin{figure}[t!]
\begin{centering}
\includegraphics[width=0.24\textwidth]{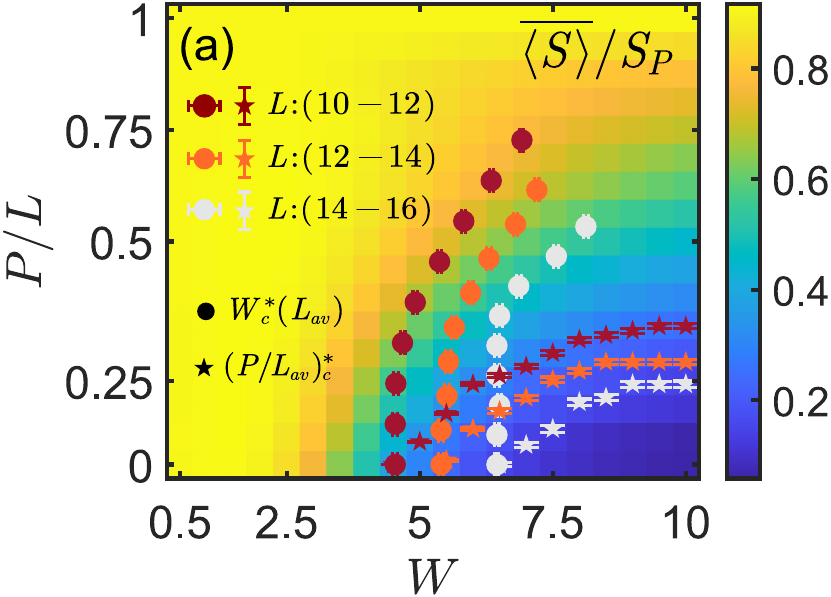}
\includegraphics[width=0.235\textwidth]{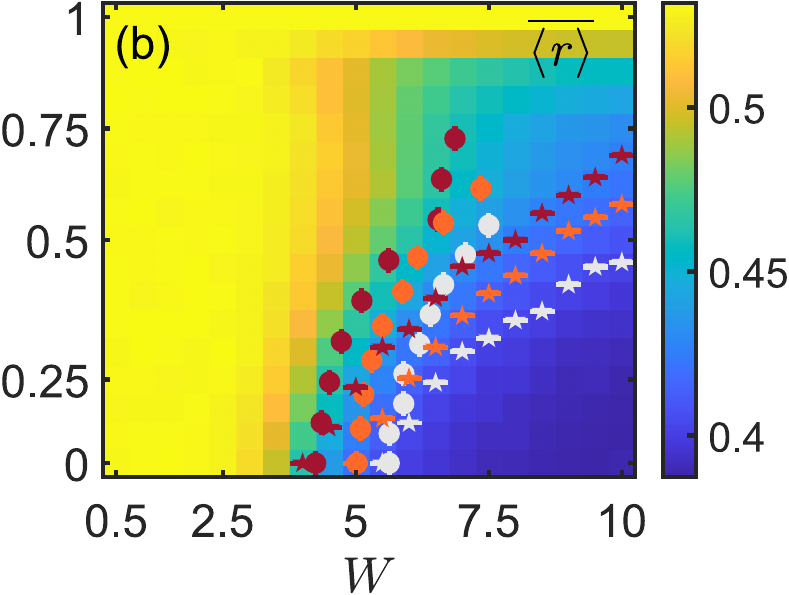}
\end{centering}
\caption{(a) averaged EE $\overline{\langle S\rangle}{/}S_{P}$ and (b) averaged GR $\overline{\langle r \rangle}$ as a function of $W$ and $P{/}L$ in a system of size $L{=}16$. In both panels, the solid circles represent the estimated $W_c^{*}(L_{av})$ for fixed $P$'s. 
The other colorful markers represent the extracted $(P{/}L_{av})_{c}^{*}$ for fixed $W$'s. As graphically visualized in Fig.~\ref{fig:1}(b), three distinct regions are classified as $1.$ ergodic, $2.$ intermediate, and $3.$ localized. 
}
\label{fig:3}
\end{figure}

To illustrate a complete picture of the effect of both disorder strength and the weak-disorder region, in Fig.~\ref{fig:3}(a) we plot $\overline{\langle S \rangle}/S_{\mathrm{P}}$ versus $W$ and $P{/}L$, obtained in a system of size $L{=}16$.
The light yellow area represents the ergodic phase and the dark blue determines the MBL spot.
As mentioned above, there are two ways to obtain the phase boundary (i) we can fix $P$ and vary $W$; or (ii) we can fix disorder strength $W$ and vary $P$. 
In the first approach, we find the intersection of $\overline{\langle S \rangle}/S_{\mathrm{P}}$ curves
for system of sizes $L$ and $L{+}2$ when they are plotted as a function of $W$ for a given $P$, e.g. see Fig.~\ref{fig:2}(a) for $P{=}0$. The intersection point $W_c^*(L_{av})$, with $L_{av}{=}L{+}1$ is taken as the finite-size transition point, at a given $P$, see Appendix~(\ref{app.Phase_bounday}) for more details. 
The results for $W_c^*(L_{av})$ are shown in Fig.~\ref{fig:3}(a) with solid circles. Different lines correspond to different sets of system sizes that are used to extract  $W_c^*(L_{av})$. 
The smooth dependence of the EE on $L$ and $W$, especially at $P{=}0$, guarantees that $W_c^{*}(L_{av})$ is a lower bound for the MBL transition point in the thermodynamic limit, namely $W_c$.
As it is evident, by increasing  $P$, the phase boundary initially remains straight and then bends after $P{/}L{>}0.25 $. This shows that the MBL phase is indeed stable against the presence of small thermal regions. On the other hand, one can also determine the phase boundary between the ergodic and the MBL phases by fixing $W$ and varying $P$.
For every given $W$, we determine the transition point $(P{/}L_{av})_c^*$ by finding the intersection point of $\overline{\langle S \rangle}/S_{\mathrm{P}}$ curves for systems of size $L$ and $L{+}2$, when they are plotted as a function of $P/L$ for a fixed value of $W$, e.g. see Fig.~\ref{fig:2_second_part}(a) for $W{=}9$.
The results are shown by stars in  Fig.~\ref{fig:3}(a). 
Interestingly, the phase boundary between the MBL and the ergodic phase is found to be different depending on the two above approaches. This suggests that there exists an intermediate region between the ergodic and the MBL phase, which we cannot decisively label as MBL or ergodic. We call this area in the phase diagram the intermediate phase, which is also schematically shown in Fig.~\ref{fig:1}(b). As discussed for Figs.~\ref{fig:2_second_part}(a) and (d) (see also Figs.~\ref{fig:S6} and \ref{fig:S7} in the Appendices for other $W$ values), this intermediate phase tends to vanish in the thermodynamic limit as the localization to ergodic transition accelerates by increasing the system size.  

Having elucidated the EE behavior for inspecting the stability of MBL, we now focus on the GR.
In Fig.~\ref{fig:3}(b) we report the obtained $\overline{\langle r \rangle}$ as a function of $W$ and $P{/}L$ for a system of size $L{=}16$. By repeating the above analysis to determine the phase boundaries, one again finds different phase boundaries depending on whether $W$ is varied while $P/L$ is kept fixed or $P/L$ is varied while $W$ is fixed. Similar to the EE, the GR also shows relative stability for the MBL phase for small values of $P/L$, which is evident in the bending of the phase boundary as $P/L$ increases. Clearly, the GR analysis also leaves an intermediate phase between the ergodic and the MBL phases, which is not fully decisive. Therefore, both EE and GR predict an intermediate phase for finite systems which lies between the MBL and the ergodic phases.

\section{Dynamical analysis} \label{Dynamics}
The impact of thermal region $\mathcal{P}$ on MBL instability should also be observable in the dynamics of the system. We consider the time evolution of EE when the system is initially prepared in the N\'eel state $\ket{\psi(0)} {=} \ket{\uparrow  \downarrow  {\cdots}  \uparrow \downarrow}$. The quantum state at time $t$ is then given by $\ket{\psi(t)} {= }e^{-itH} \ket{\psi(0)}$, whose EE is computed by tracing out the spins in half of the system. In numerical calculations, we employ exact diagonalization to obtain the evolved state and use  $200{-}400$ disorder realizations (depending on $L$) to get average EE $\overline{S(t)}$. It is well-known that in the ergodic phase EE grows linearly in time, namely $\overline{S (t)}/S_{\mathrm{P}}{\sim}t$, till 
it saturates to the volume-law value and, therefore, results in $\overline{S (t)}/S_{\mathrm{P}}{\rightarrow}1$~\cite{nakagawa2018universality}. On the other hand, in the MBL phase, EE grows logarithmically in time, namely $\overline{S (t)}/S_{\mathrm{P}}{\sim}\ln(t)$~\cite{EGSR3}, and eventually saturates into its value for an entangled state which may exhibit extensive entanglement, though lower than thermal states~\cite{Rev3,EGSR3}. 

\begin{figure}[t!]
\begin{centering}
\includegraphics[width=0.23\textwidth]{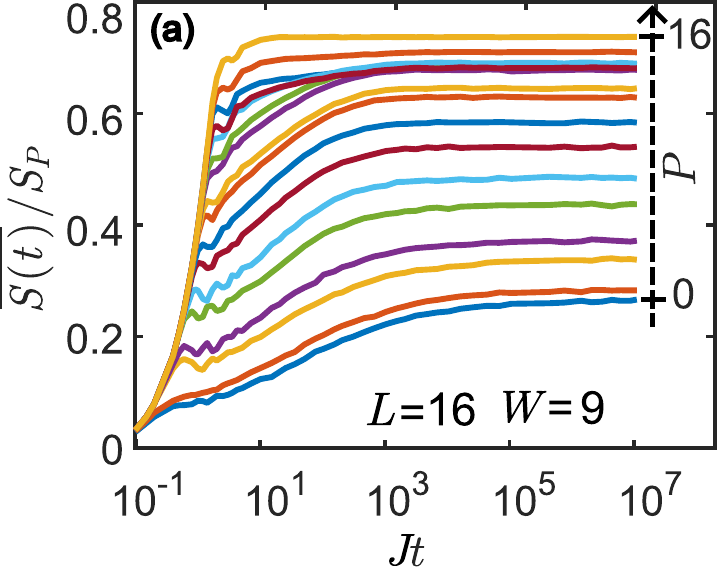} 
\includegraphics[width=0.23\textwidth]{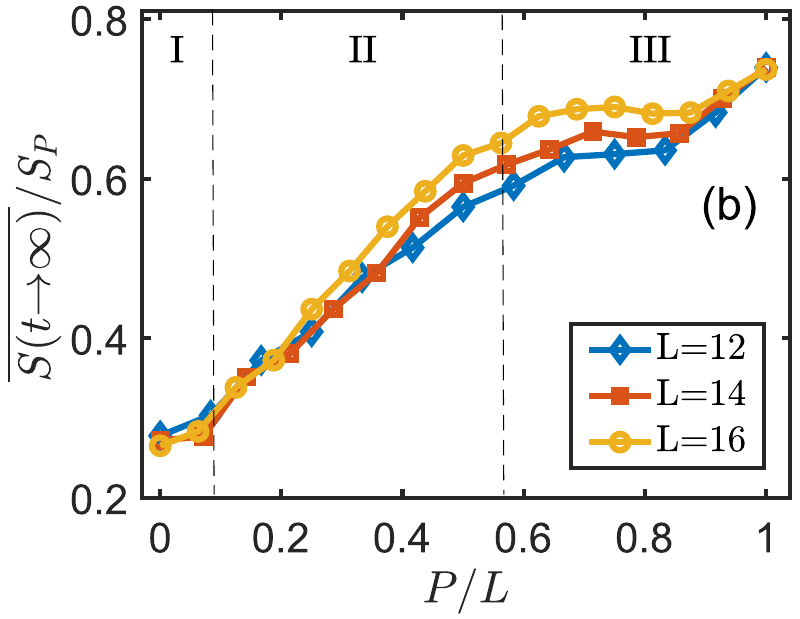}
\end{centering}
\caption{(a) The dynamic of EE in a system of size $L{=}16$ for different values of $P$, when the system with $P{=}0$ is determined as localized in the presence of strong disorder $W{=}9$.
(b) the saturated values of the EE, namely $\overline{S(t{\rightarrow\infty)}}{/}S_{P}$, as a function of $P{/}L$ when the size of the system varies and $W{=}9$. The results are obtained using $200{-}400$ realizations.}
\label{fig:4}
\end{figure}

In Fig.~\ref{fig:4}(a), we plot $\overline{S (t)}/S_{\mathrm{P}}$ as a function of time in a system with disorder strength $W{=}9$ for different choices of $P$. As the results show, for $P{=}0$  where the system is in the MBL  phase, EE grows logarithmically in time before saturation to $\overline{S (t{\rightarrow} \infty)}/S_{\mathrm{P}}\simeq 0.26$. In contrast, for $P{=}L$, i.e. ergodic phase, EE exhibits linear growth in time which eventually saturates to $\overline{S (t{\rightarrow} \infty)}/S_{\mathrm{P}}\simeq 0.74$. To see the impact of the thermal region $\mathcal{P}$, we focus on the saturated value of EE  at long time scales. In Fig.~\ref{fig:4}(b), we plot $\overline{S(t{\rightarrow}\infty)}/S_{\mathrm{P}}$
as a function of $P{/}L$ for $W{=}9$ and various system sizes.
Consistent with the static analysis, three distinct regions can be recognized: (i) the early plateau in region I, i.e. $P/L{\leq}0.1$, indicates a robust MBL phase which resists against thermalization; (ii) the linear rise of the saturated EE in region II, i.e. $0.1{\leq} P/L{\leq} 0.6$, marks the intermediate phase, where the system shows partial thermalization without fully reaching the volume law; and (iii) the final region III, i.e. $P/L{>} 0.6$ where $\overline{S(t{\rightarrow}\infty)}/S_{\mathrm{P}}$ remains almost flat indicating the thermal phase.
In region II, the linear increase of $\overline{S(t{\rightarrow}\infty)}/S_{\mathrm{P}}$ becomes sharper in systems of larger sizes, suggesting that the intermediate phase vanishes in the thermodynamic limit.
These observations are completely in line with our static results.

\section{Conclusion} \label{concl.}
The delicate balance between disorder strength and the minimum size of a 
thermalizing region for triggering the avalanche effect remains a central open question in understanding MBL stability. Here, we systematically address this question in an isolated Heisenberg spin chain subjected to random fields by analyzing EE and GR in both static and dynamical scenarios. Two main findings can be highlighted. First, we show that in isolated systems in order to destabilize the MBL  phase, the size of the thermal region should increase with system size such that $P/L$ exceeds a threshold value,  which our simulations suggest to be $P/L{>}0.1$.  Second, under strong disorder, enlarging the thermal region drives the system into an intermediate phase before fully thermalizing it. The intermediate phase exhibits hybrid characteristics, neither fully localized nor completely ergodic, and can be identified in both static and dynamical analysis in finite systems. The intermediate phase seems to shrink as the system size increases, suggesting that it vanishes in the thermodynamic limit.

\begin{acknowledgements}
AB acknowledges support from the National Natural Science Foundation of China (grants No. 12274059, No. 12574528 and No. 1251101297).
\end{acknowledgements}
 
\appendix 
\section{Finite-size scaling analysis} \label{app.Pyfssa}
%
%
%
%
\begin{figure*}
\begin{centering}
\includegraphics[width=0.245\textwidth]{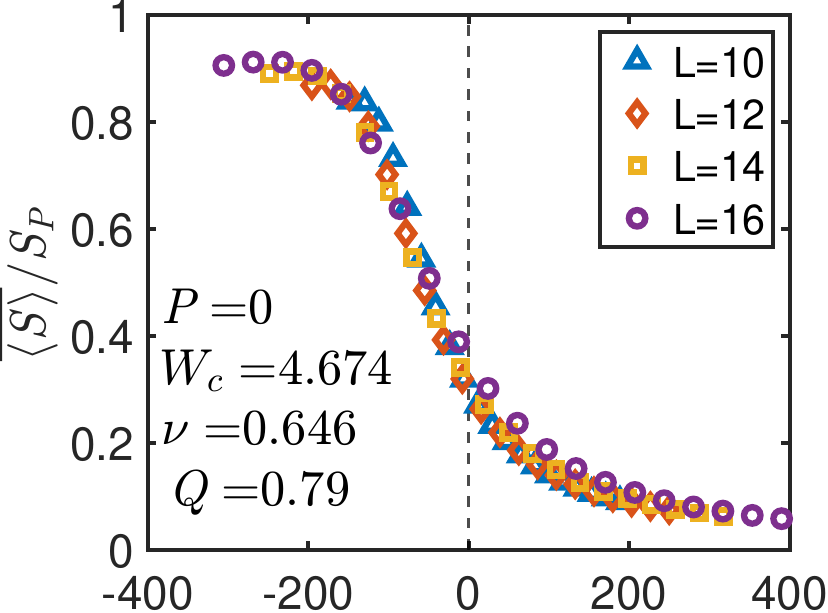} \hspace{0.0mm}
\includegraphics[width=0.23\textwidth]{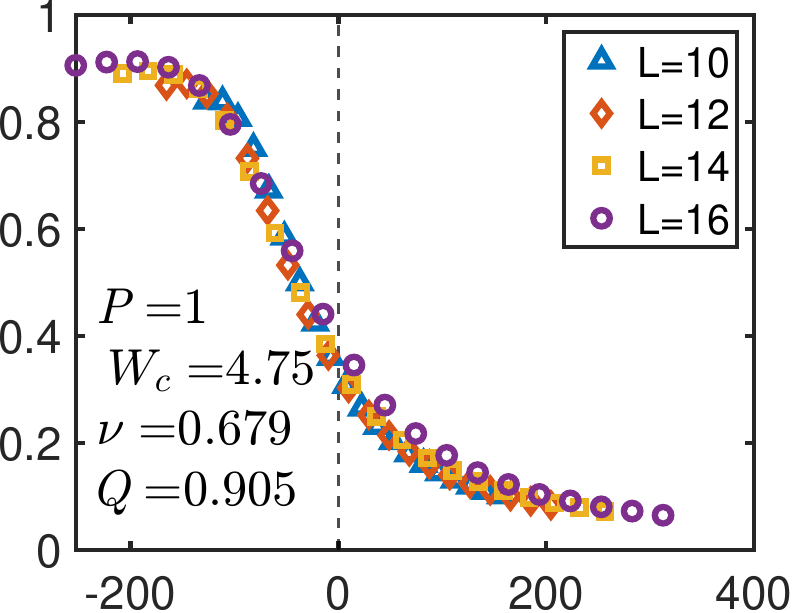}
\hspace{0.0mm}
\includegraphics[width=0.23\textwidth]{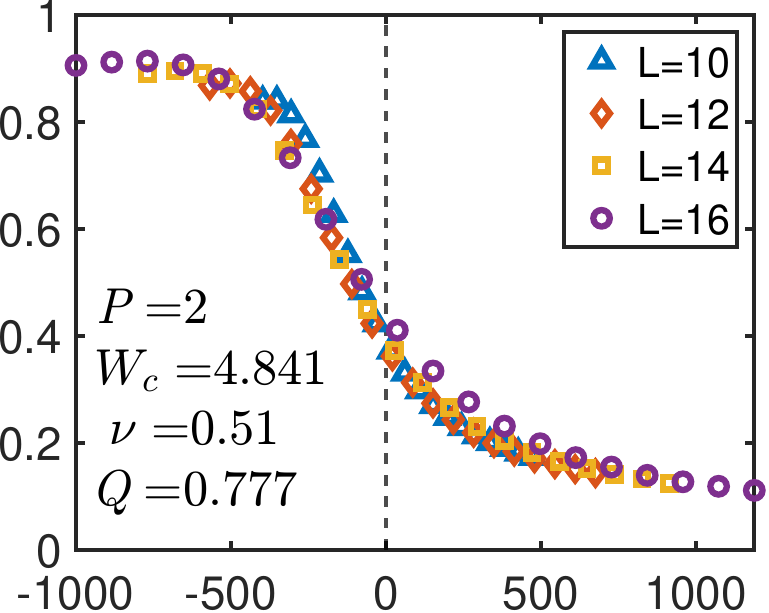}
\hspace{0.0mm}
\includegraphics[width=0.23\textwidth]{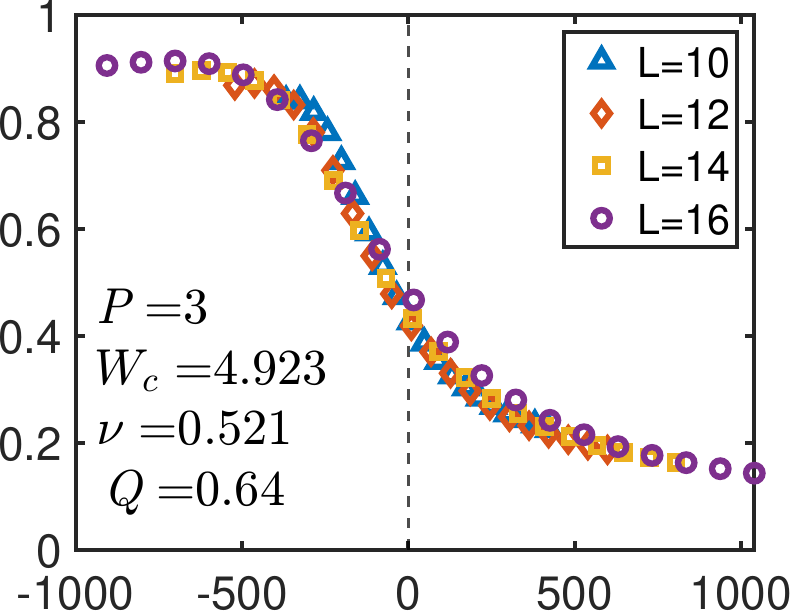} \\
 \vspace{2mm}
 \includegraphics[width=0.245\textwidth]{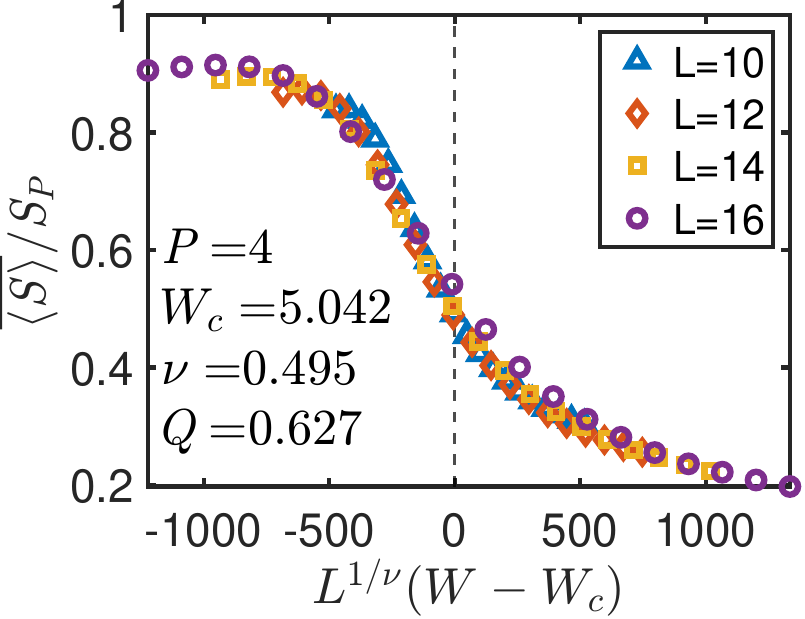} \hspace{0.0mm}
\includegraphics[width=0.23\textwidth]{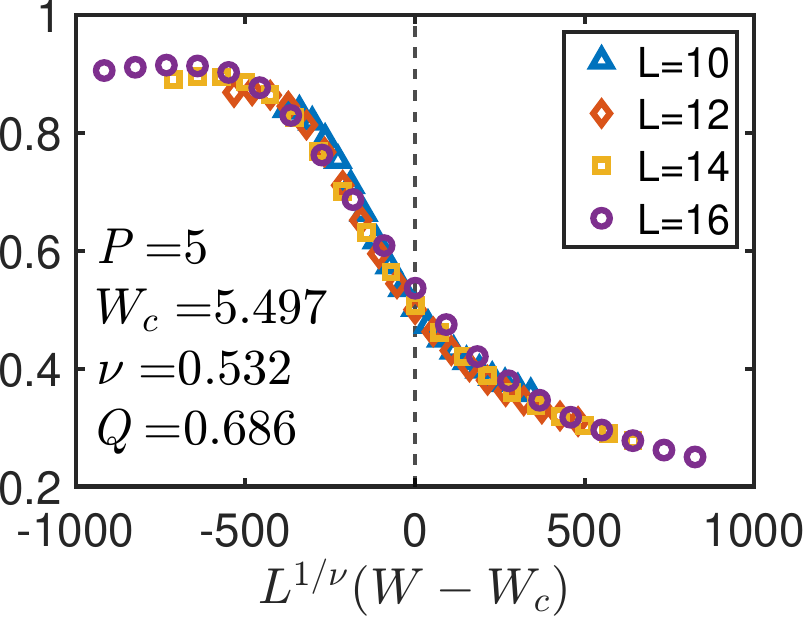}
\hspace{0.0mm}
\includegraphics[width=0.23\textwidth]{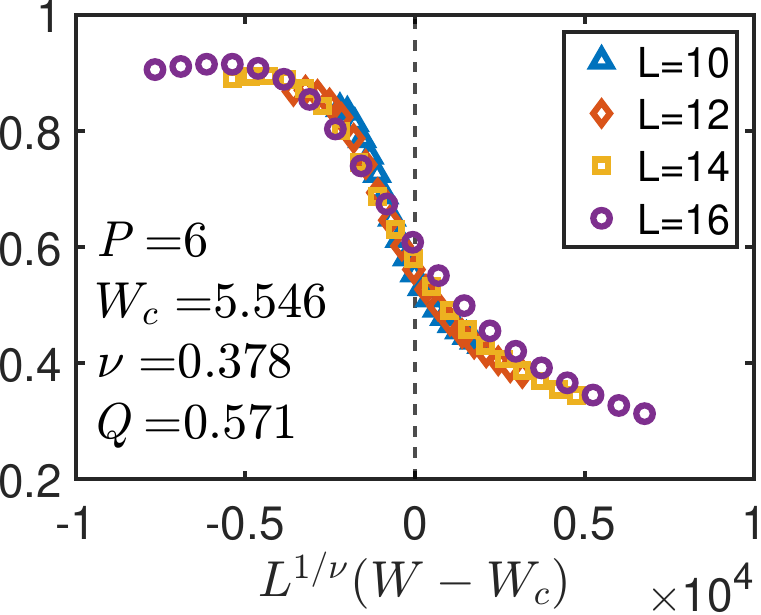}
\hspace{0.0mm}
\includegraphics[width=0.23\textwidth]{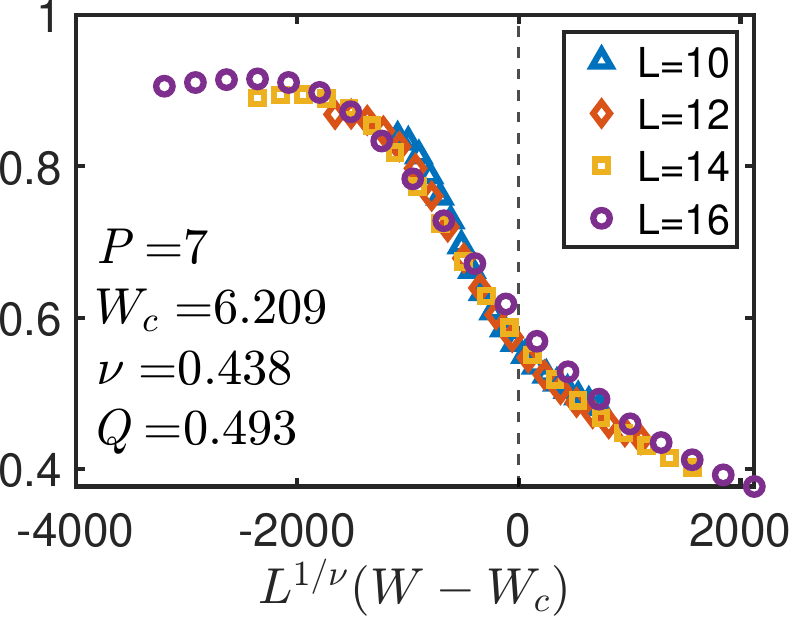} 
\end{centering}
\caption{Data collapse of the normalized EE using finite-size scaling analysis for various values of $P$. The values of the critical disorder strength, critical exponent, and the optimal value of the quality function in each case are indicated in the insets of the corresponding plot.}
\label{fig:S1}
\end{figure*}
%
%
%
%
%
%
%
%
%
\begin{figure*}
\begin{centering}
\includegraphics[width=0.24\textwidth]{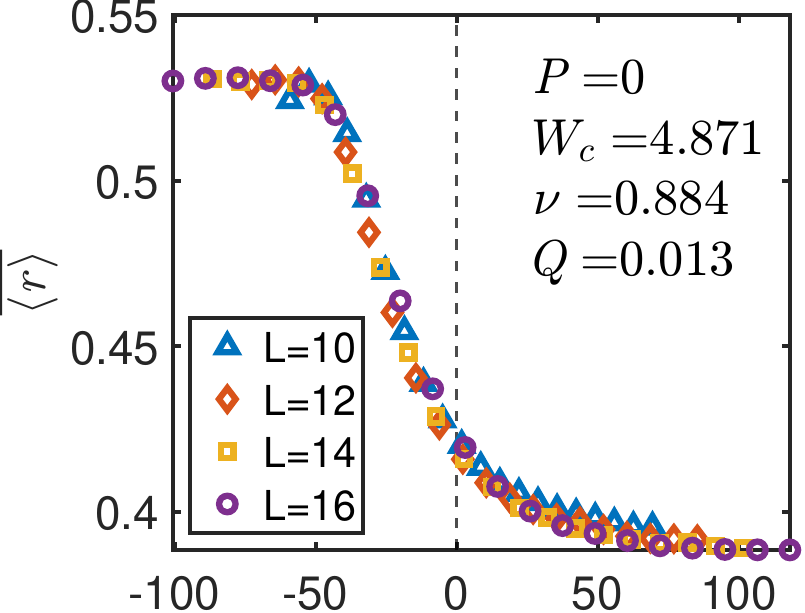} \hspace{0.0mm}
\includegraphics[width=0.23\textwidth]{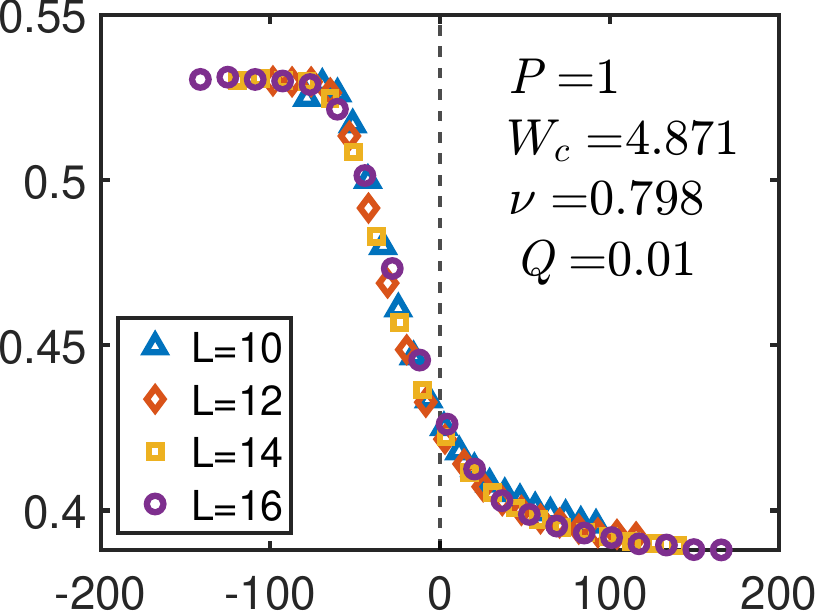}
\hspace{0.0mm}
\includegraphics[width=0.23\textwidth]{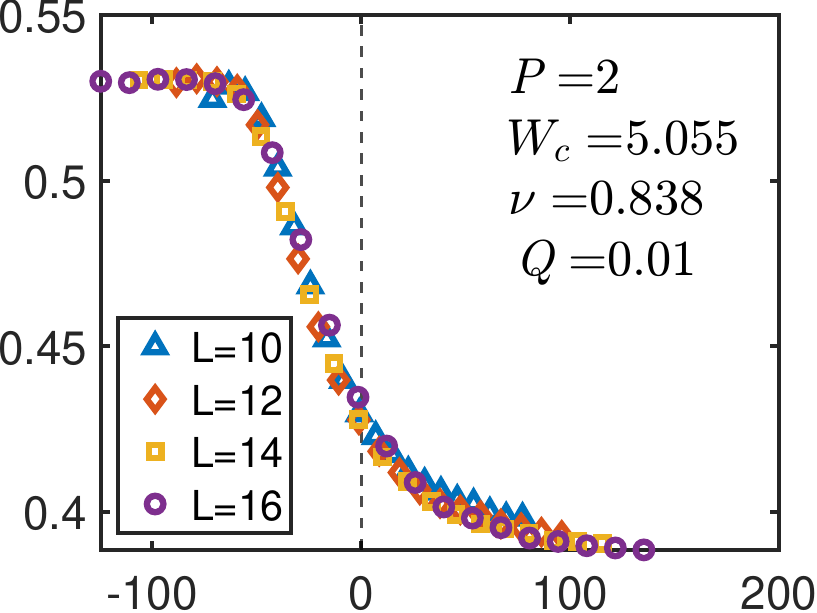}
\hspace{0.0mm}
\includegraphics[width=0.23\textwidth]{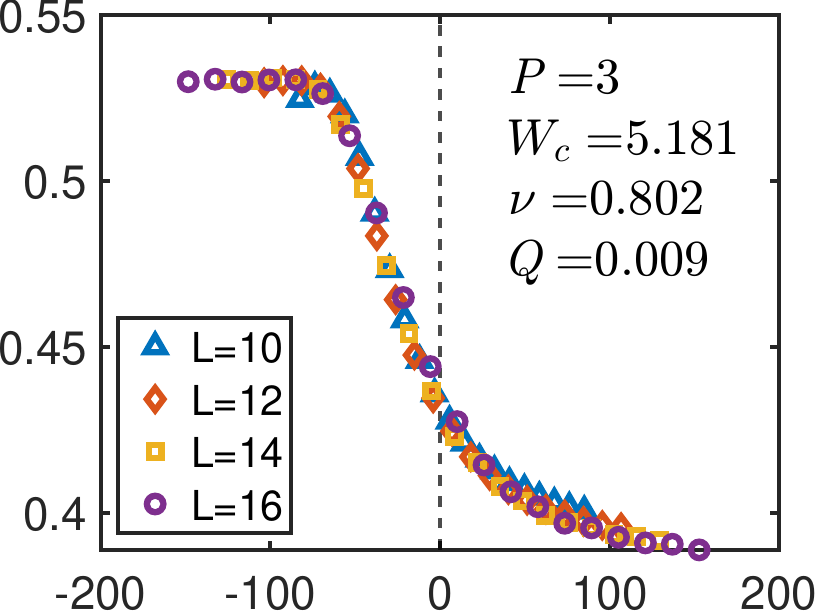} \\
 \vspace{2mm}
 \includegraphics[width=0.24\textwidth]{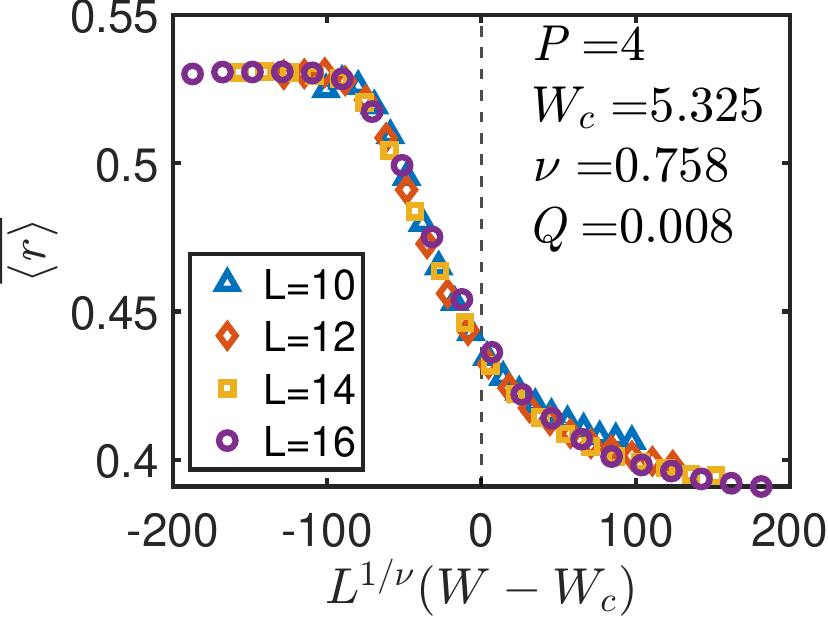} \hspace{0.0mm}
\includegraphics[width=0.23\textwidth]{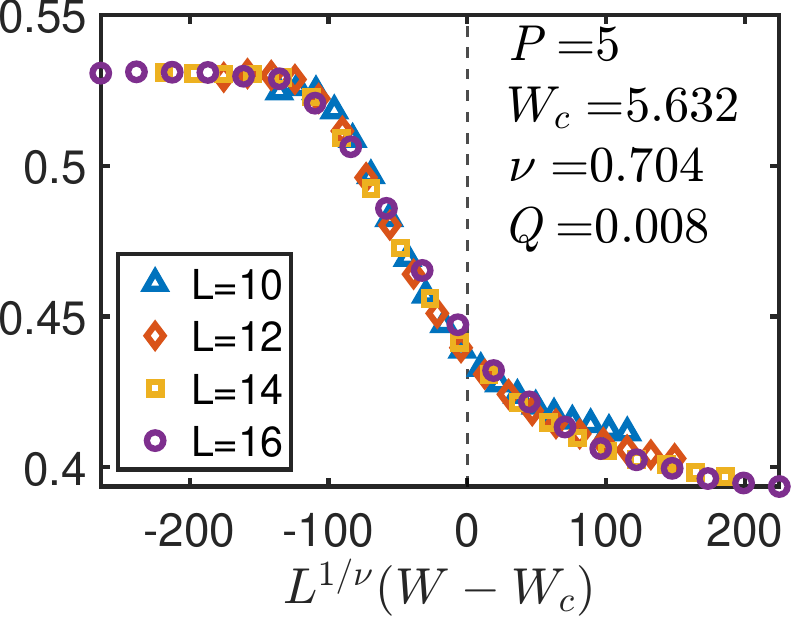}
\hspace{0.0mm}
\includegraphics[width=0.23\textwidth]{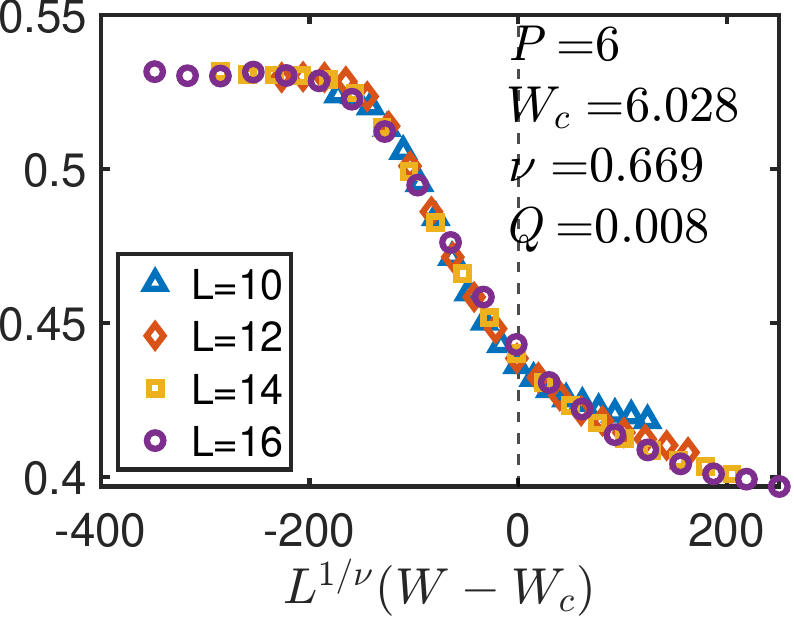}
\hspace{0.0mm}
\includegraphics[width=0.23\textwidth]{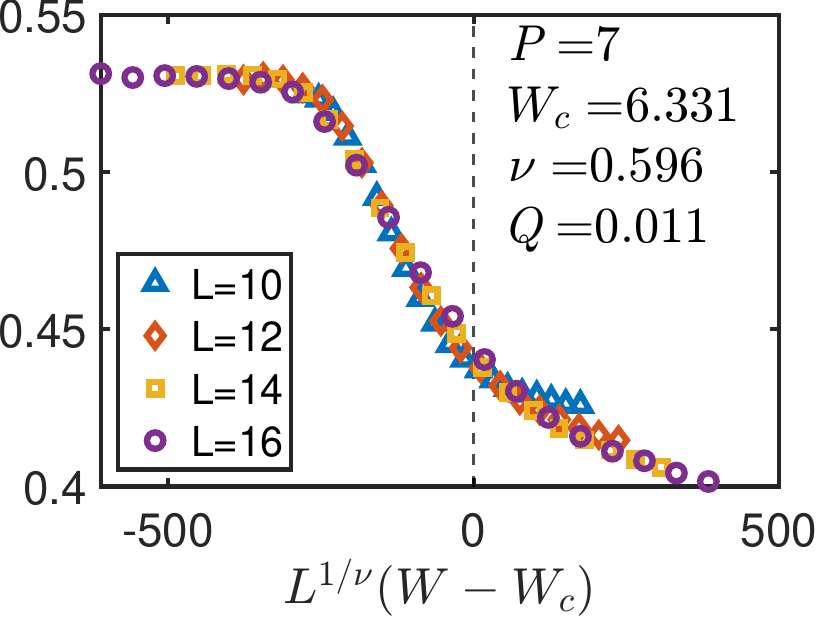}
\end{centering}
\caption{Data collapse of the averaged GR using finite-size scaling analysis for various values of $P$. The values of the critical disorder strength, critical exponent, and the optimal value of the quality function in each case are indicated in the insets of the corresponding plot.}
\label{fig:S2}
\end{figure*}
%
%
%
%
\begin{figure*}
\begin{centering}
\includegraphics[width=0.28\textwidth]{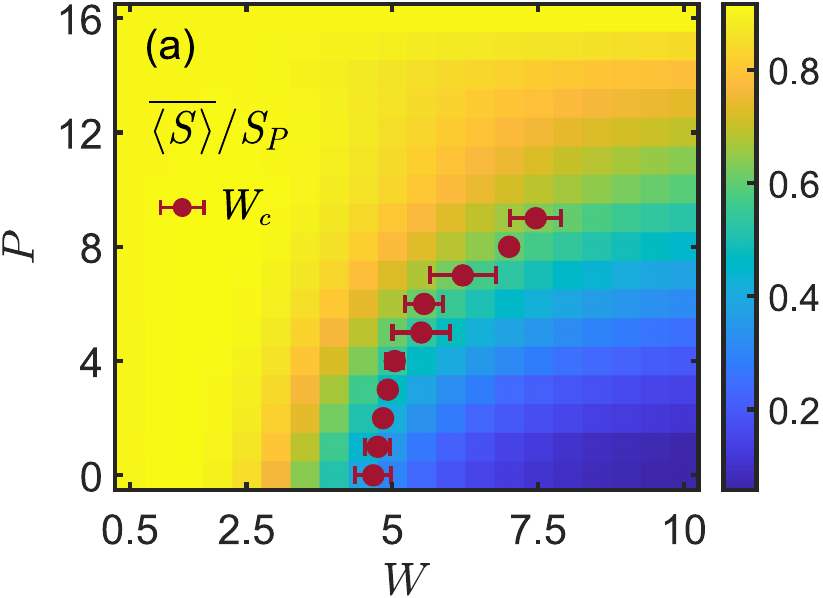} \hspace{0.0mm}
\includegraphics[width=0.28\textwidth]{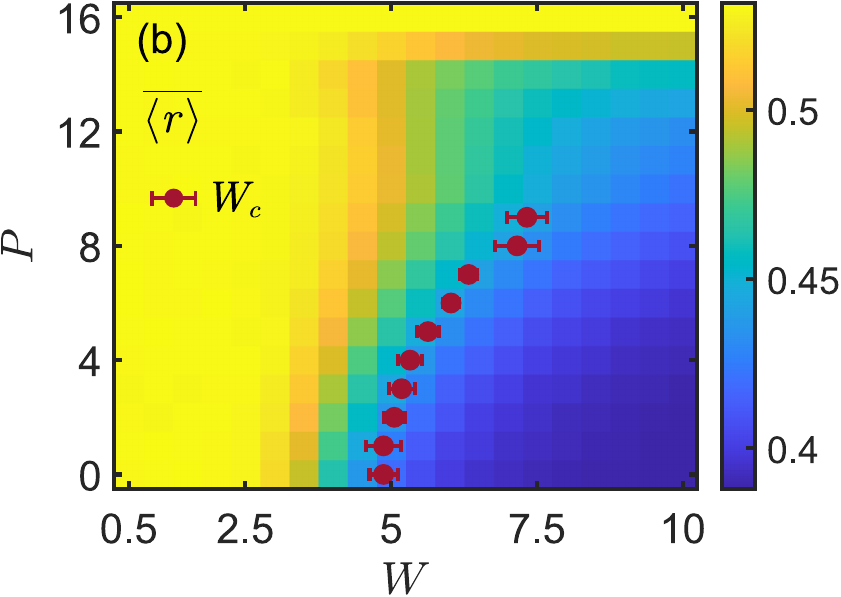}
\end{centering}
\caption{(a) averaged EE and (b) averaged GR computed for the system size $L{=}16$ as a function of $W$ and $P$. In both panels, the overlaid dots represent the critical disorder strength $W_c$ estimated for $P{=}0,1,{\cdots},9$ using finite-size scaling analysis for system sizes $L{=}10,12,14, \ \text{and} \ 16$.}
\label{fig:S3}
\end{figure*}
We consider the MBL transition as a second-order quantum phase transition and use finite-size scaling analysis to extract the critical disorder strength $W_c$ as a function of $P$ (the size of the low disorder region).
For a quantity $O_L(W)$ that depends on the system size $L$ and disorder strength $W$, we use the following finite-size scaling ansatz:
\begin{align}\label{eq:_S1}
\overline{\langle O_L (W) \rangle} = f(L^{1/\nu} (W-W_c)),
\end{align}
here, $\nu$ is the universal critical exponent for the second-order phase transition, and $f(\cdot)$ is a dimensionless function whose specific form is unknown.
To estimate the optimal values of the critical parameters $W_c$ and $\nu$ for our finite-size data, we use the ansatz given by Eq.~(\ref{eq:_S1}) and plot our quantity of interest as a function of the scaled disorder strength, i.e. $x{=}L^{1/\nu} (W-W_c)$, for different system sizes $L$.
According to Eq.~(\ref{eq:_S1}), these curves collapse to each other if $W_c$ and $\nu$ are chosen properly.
We use PYTHON package PYFSSA for finite-size scaling analysis~\cite{pyfssa,Melchert2004} that optimizes a quality function $Q$ as the quality of data collapse and estimates the optimal values of the scaling parameters. 
The quality function, elaborated in~\cite{HoudayerPRB}, for the quantity $O$ can be written as,
\begin{equation}\label{eq:_S3}
Q_O  = \frac{1}{T} \sum_{i,j}\frac{(O_{ij} - \mathcal{O}_{ij})^2}{dO_{ij}^2 + d\mathcal{O}_{ij}^2}.
\end{equation}
Here, $O_{ij}$ represents the values of the quantity $O$ and $dO_{ij}$ represents the corresponding standard deviations at the scaled disorder strength $x_{ij}{=}L_{i}^{1/\nu} (W_j-W_C)$.
Similarly, $\mathcal{O}_{ij}$ and $d\mathcal{O}_{ij}$ respectively represent the estimated values of the master function $f(\cdot)$ and the corresponding standard deviation at $x_{ij}$.
The normalization factor $T$ represents the number of terms for which $\mathcal{O}_{ij}$ and $d\mathcal{O}_{ij}$ are defined. The sum in Eq.~(\ref{eq:_S3}) includes only those $T$ number of terms. 
In the case of an optimal data collapse, the quality function is minimized, i.e., $Q_O \sim 1$. 

In Figs.~\ref{fig:S1} and~\ref{fig:S2}, we show the obtained data collapse for the normalized entanglement entropy (EE) and the gap ratio (GR) for various values of $P$, respectively. The optimized values of the critical parameters ($W_c$ and $\nu$) and the quality function $Q$ are reported, respectively. 
In the absence of a low-disorder region, for EE, we obtain $W_c = 4.674 \pm 0.308$, $\nu = 0.646$, and $Q = 0.79$.
Regarding GR, we obtain $W_c = 4.871 \pm 0.243$, $\nu = 0.884$, and $Q = 0.013$.
In Figs.~\ref{fig:S3}(a) and (b), we show the critical disorder strength $W_c$ as a function of $P$ in the phase diagram for the EE and the GR, respectively.
Both panels show that $W_c$ remains almost unchanged for the first few values of $P$ and then increases by enlarging the thermal region $\mathcal{P}$.

\begin{figure*}[t]
\begin{centering}
\includegraphics[width=0.235\textwidth]{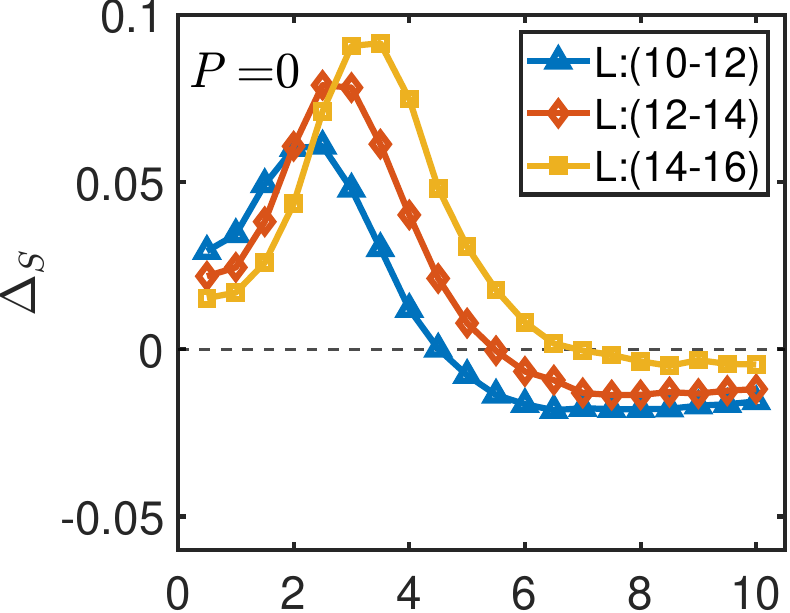} \hspace{0.0mm}
\includegraphics[width=0.23\textwidth]{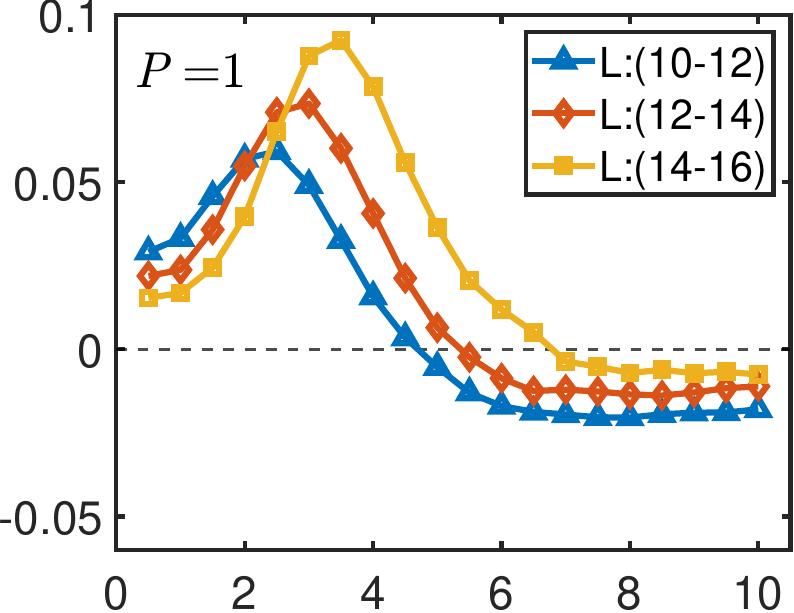}
\hspace{0.0mm}
\includegraphics[width=0.23\textwidth]{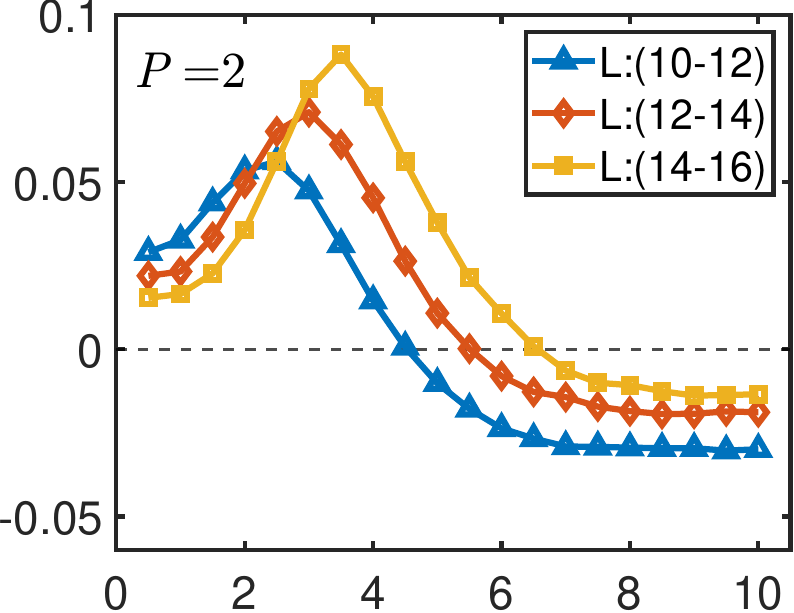}
 \hspace{0.0mm}
 \includegraphics[width=0.23\textwidth]{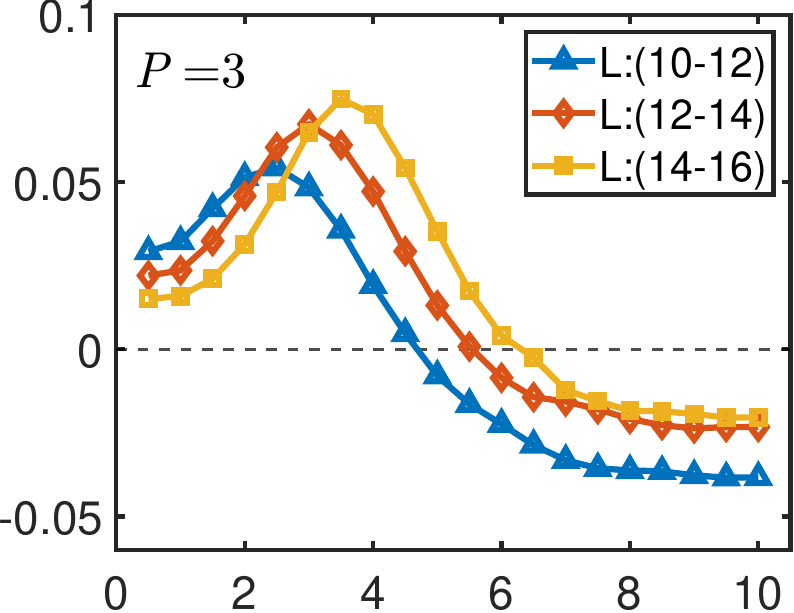} 
 \hspace{0.0mm}
\includegraphics[width=0.235\textwidth]{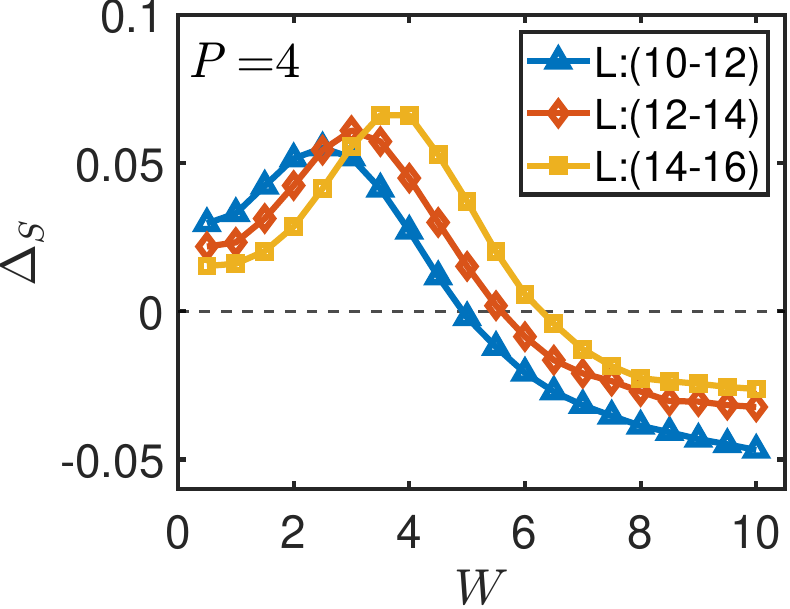}
\hspace{0.0mm}
\includegraphics[width=0.23\textwidth]{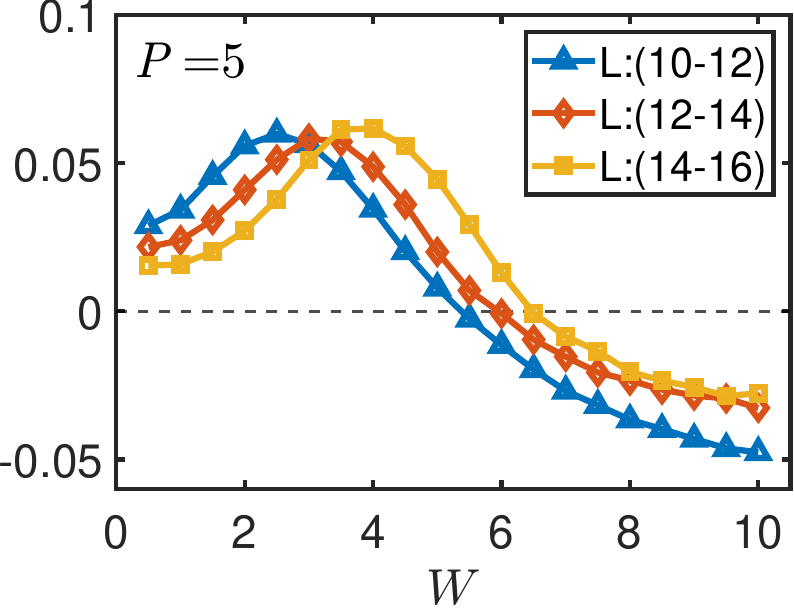}
\hspace{0.0mm}
 \includegraphics[width=0.23\textwidth]{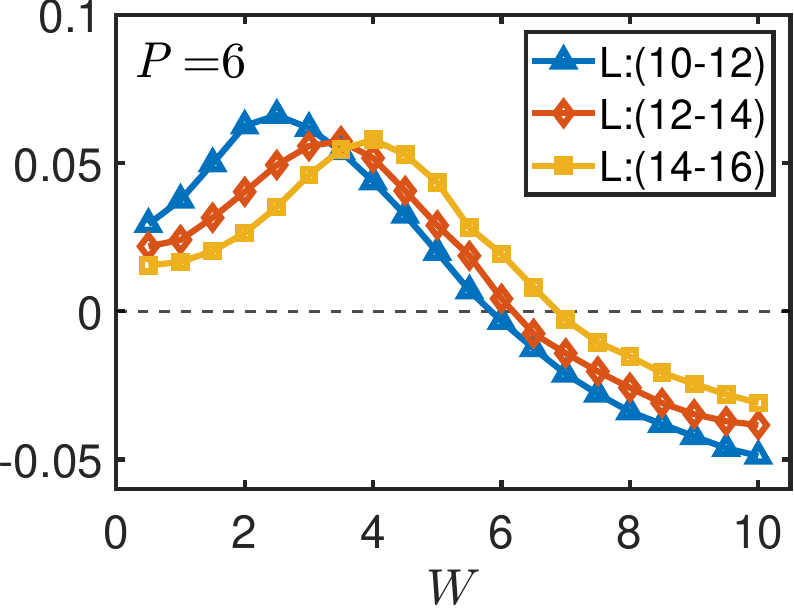} \hspace{0.0mm}
\includegraphics[width=0.23\textwidth]{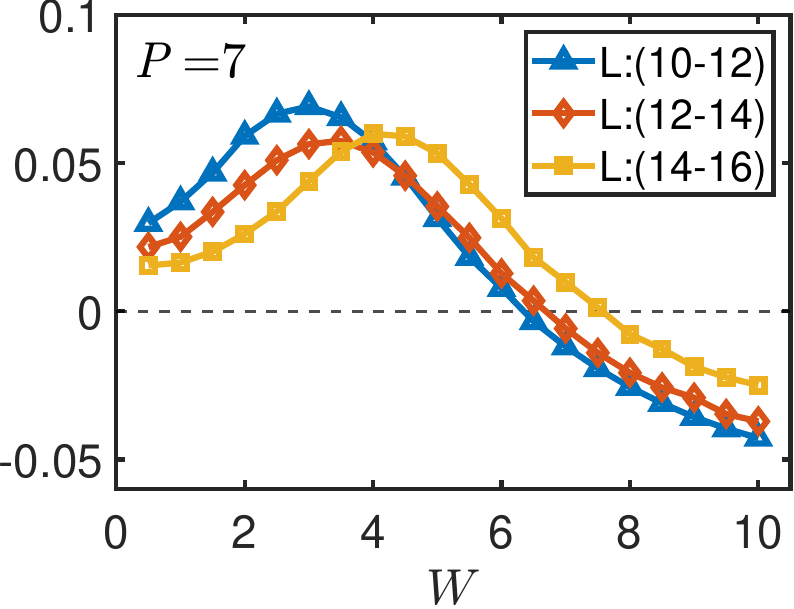}
\end{centering}
\caption{Difference of the normalized EE for all pairs of $L$ and $L+2$ plotted as a function of the disorder strength for various values of $P$. For a given $L_{av}$, the critical disorder $W_c^*(L_{av})$ is estimated from the crossing of the corresponding curve with the zero line (indicated by the black dashed line). For a fixed $P$, the drifts of $W_c^*(L_{av})$ from smaller to larger values (corresponding to smaller to larger $L$) is visible.}
\label{fig:S4}
\end{figure*}
%
\begin{figure*}[t]
\begin{centering}
\includegraphics[width=0.235\textwidth]{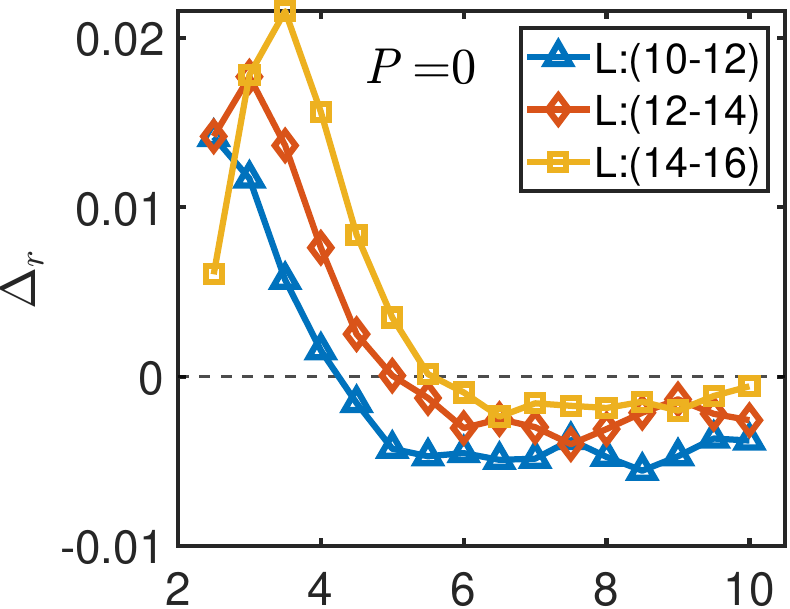} \hspace{0.0mm}
\includegraphics[width=0.23\textwidth]{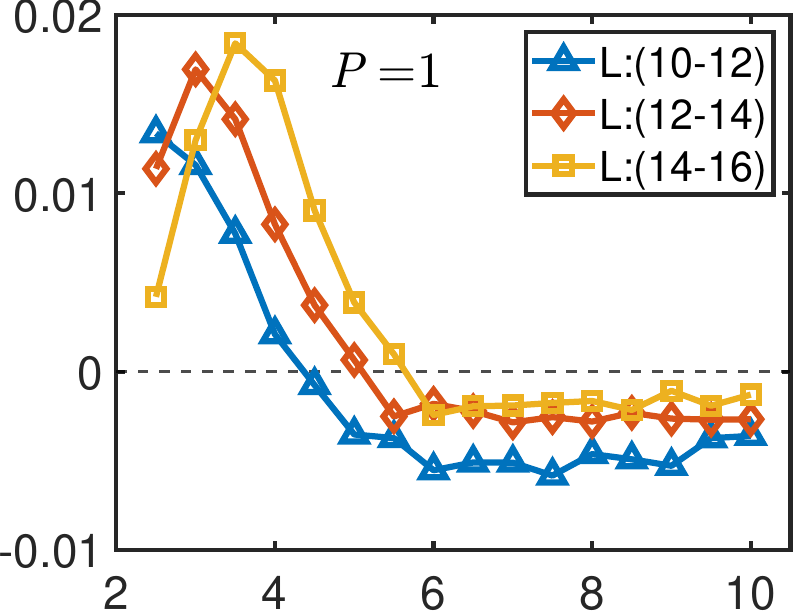}
\hspace{0.0mm}
\includegraphics[width=0.23\textwidth]{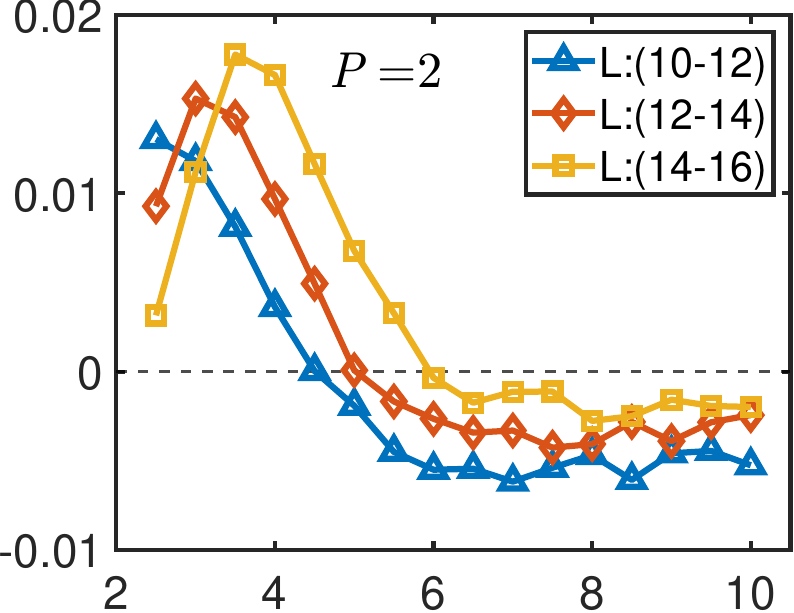}
\hspace{0.0mm}
\includegraphics[width=0.23\textwidth]{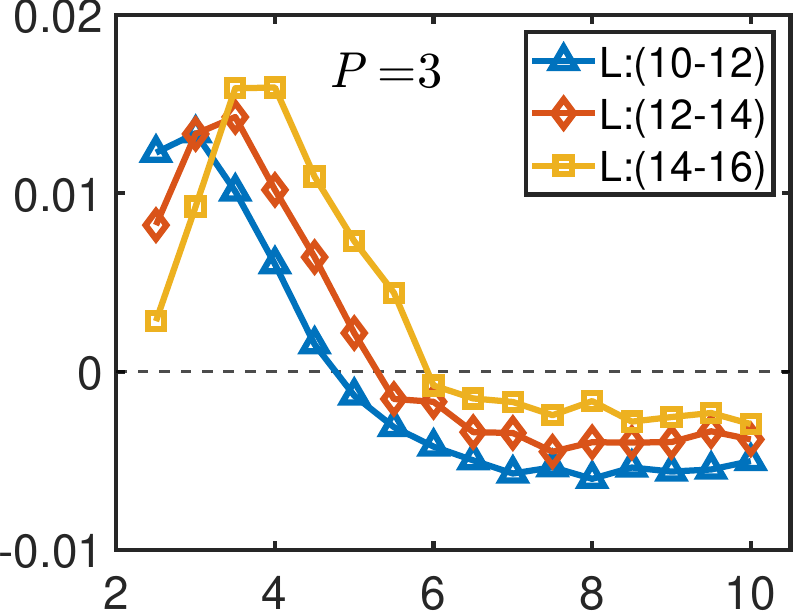} \hspace{0.0mm}
\includegraphics[width=0.235\textwidth]{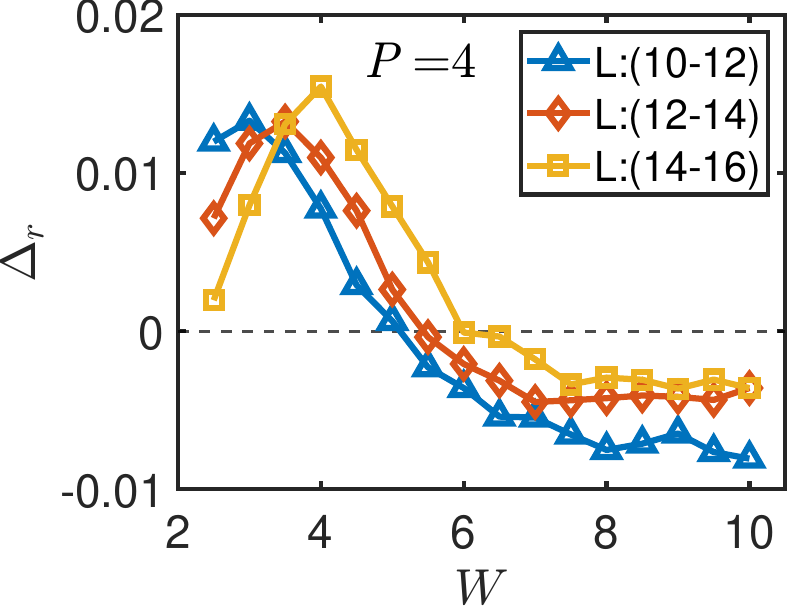}
\hspace{0.0mm}
\includegraphics[width=0.23\textwidth]{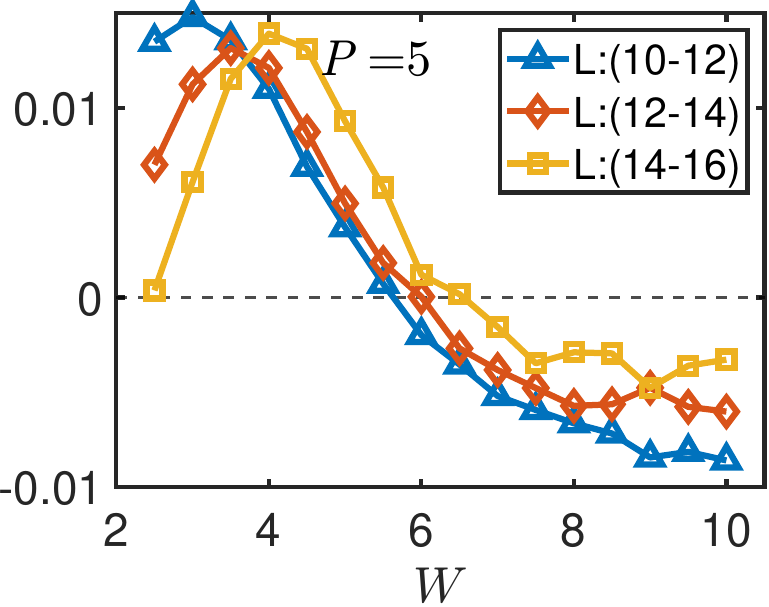}
\hspace{0.0mm}
\includegraphics[width=0.23\textwidth]{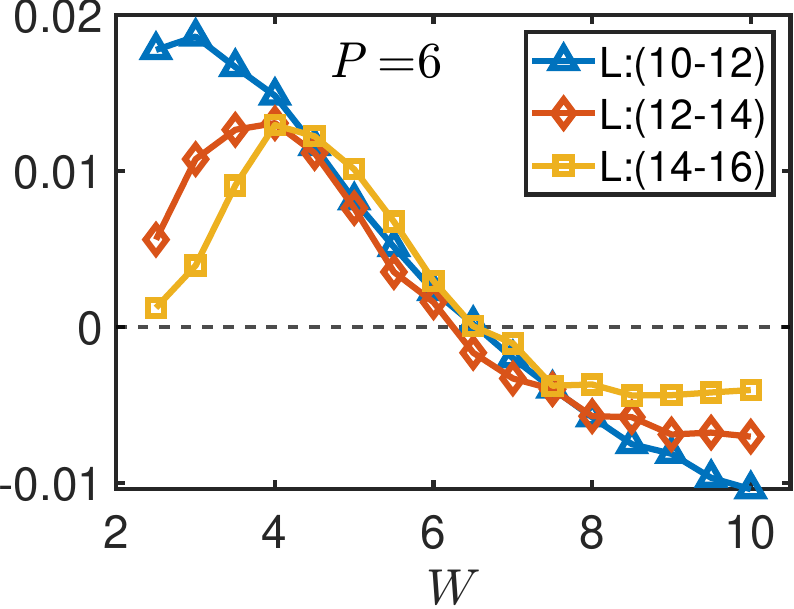}\hspace{0.0mm}
\includegraphics[width=0.23\textwidth]{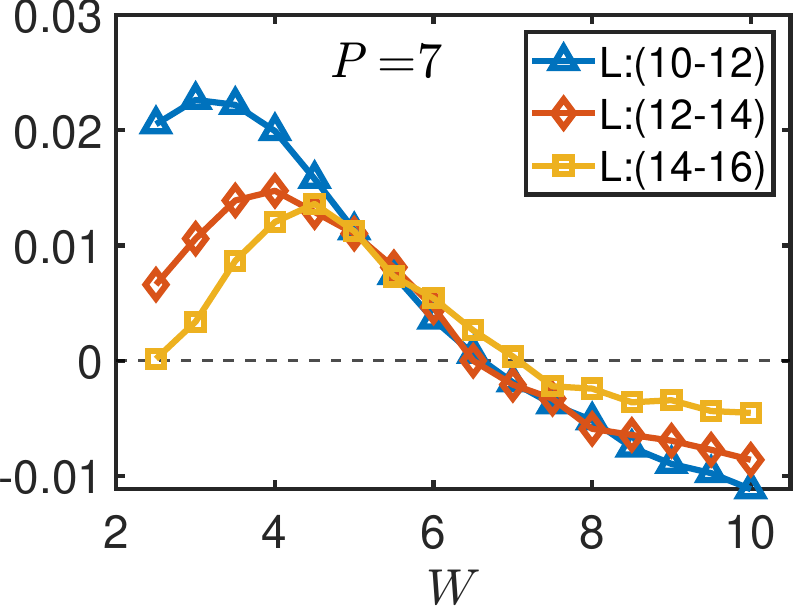}
\end{centering}
\caption{Difference of the averaged GR for all pairs of $L$ and $L+2$ plotted as a function of the disorder strength for various values of $P$. For a given $L_{av}$, the critical disorder $W_c^*(L_{av})$ is estimated from the crossing of the corresponding curve with the zero line (indicated by the black dashed line). For a fixed $P$, the drifts of $W_c^*(L_{av})$ from smaller to larger values (corresponding to smaller to larger $L$) is visible.}
\label{fig:S5}
\end{figure*}

\begin{figure*}
\begin{centering}
\includegraphics[width=0.235\textwidth]{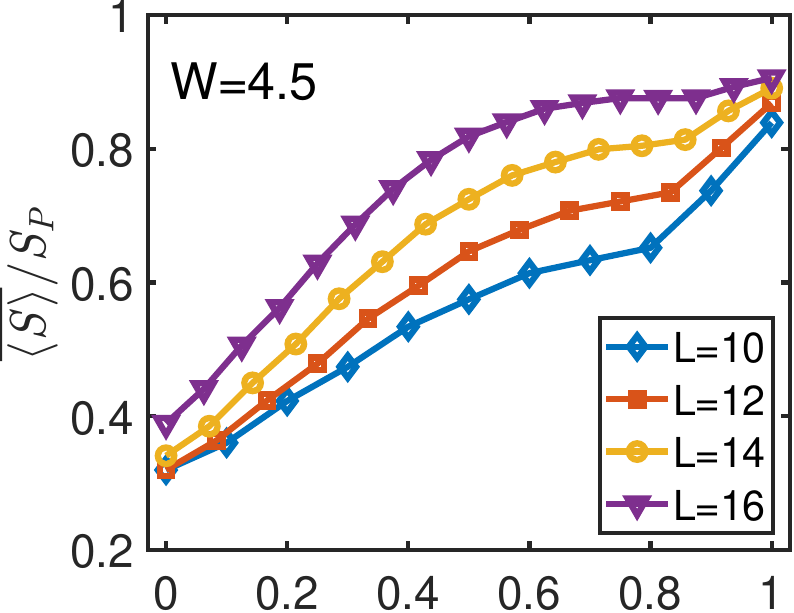} \hspace{0.0mm}
\includegraphics[width=0.225\textwidth]{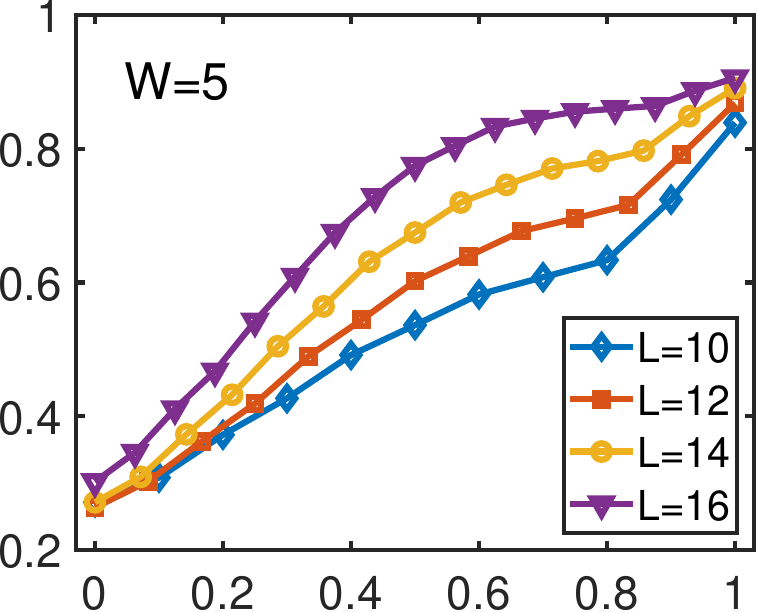}
\hspace{0.0mm}
\includegraphics[width=0.225\textwidth]{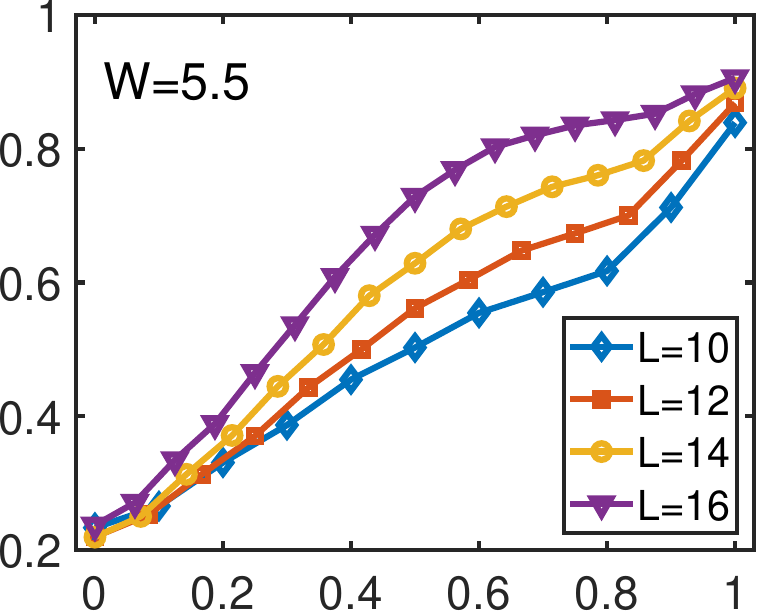}
\hspace{0.0mm}
\includegraphics[width=0.225\textwidth]{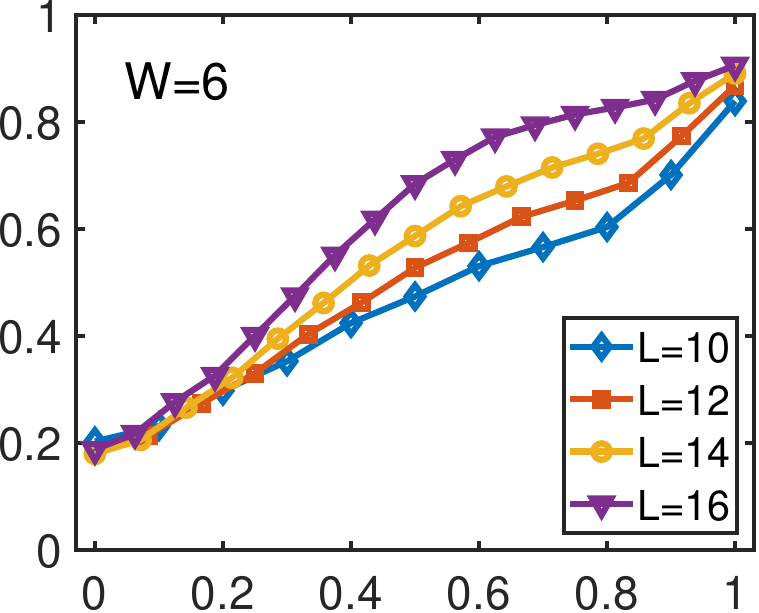} \\
\vspace{2mm}
\includegraphics[width=0.235\textwidth]{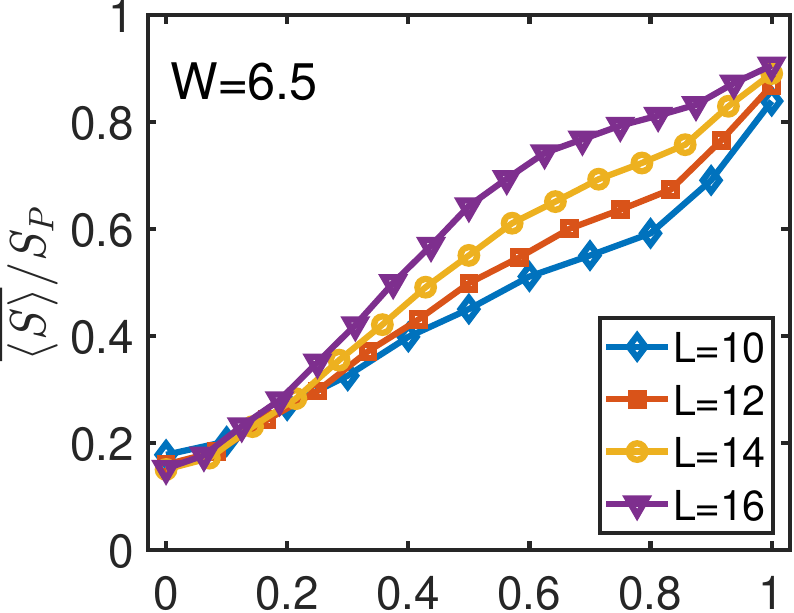} \hspace{0.0mm}
\includegraphics[width=0.225\textwidth]{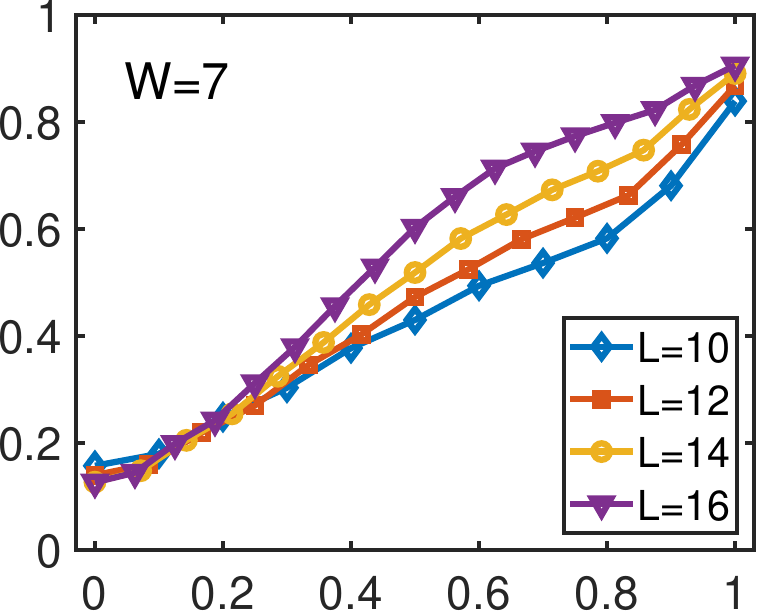}
\hspace{0.0mm}
\includegraphics[width=0.225\textwidth]{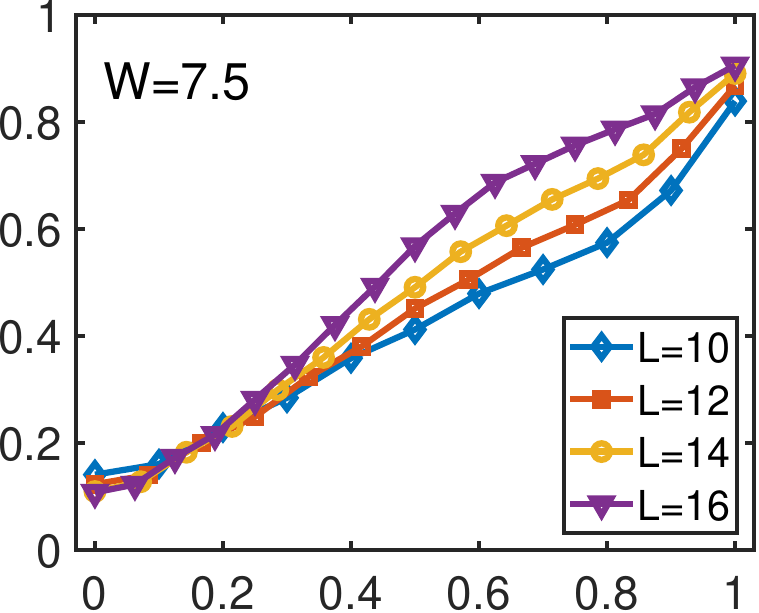}
\hspace{0.0mm}
\includegraphics[width=0.225\textwidth]{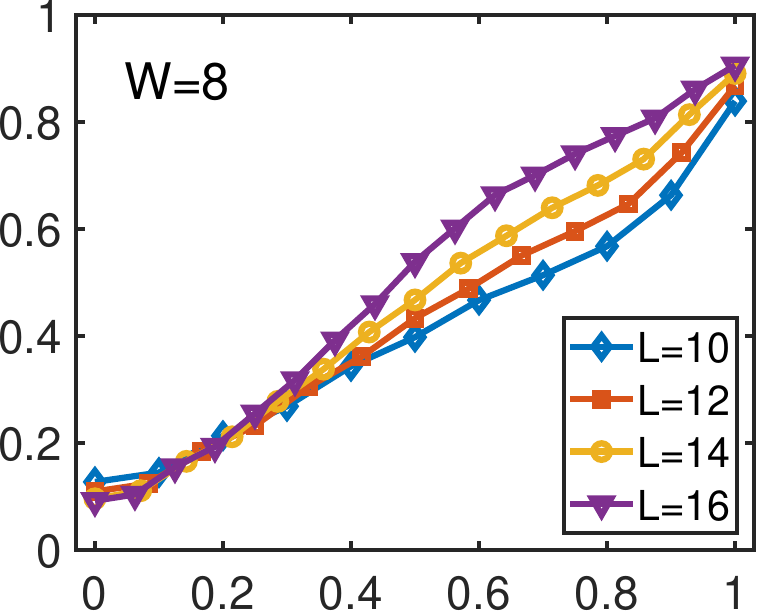} \\
\vspace{2mm}
 \includegraphics[width=0.235\textwidth]{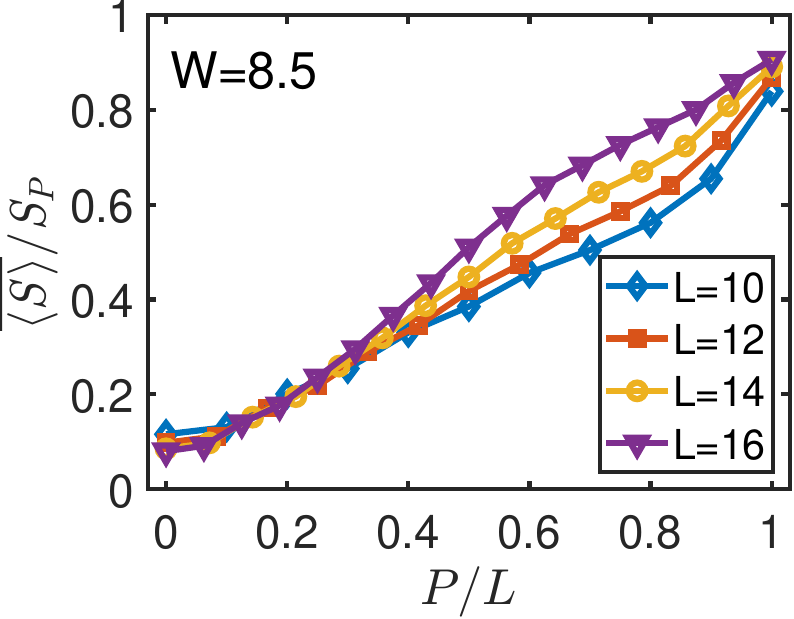} \hspace{0.0mm}
\includegraphics[width=0.225\textwidth]{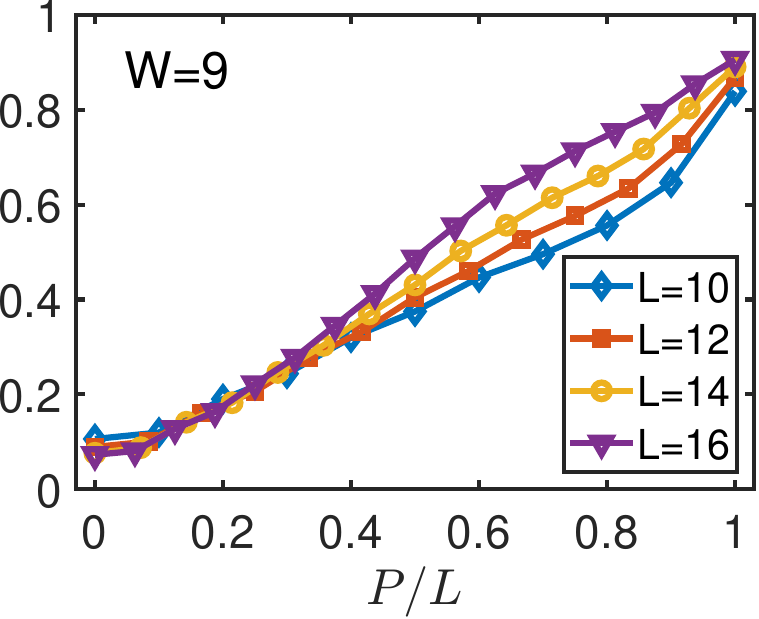}
\hspace{0.0mm}
\includegraphics[width=0.225\textwidth]{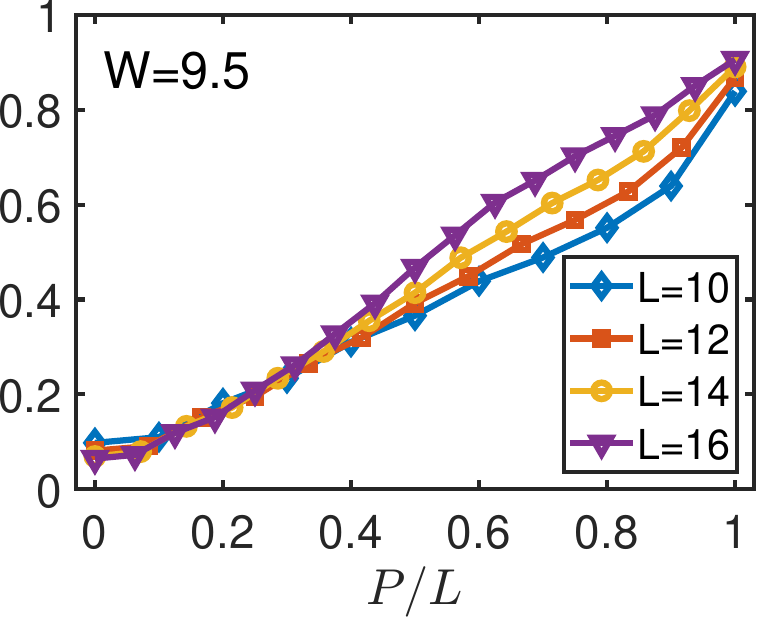}
\hspace{0.0mm}
\includegraphics[width=0.225\textwidth]{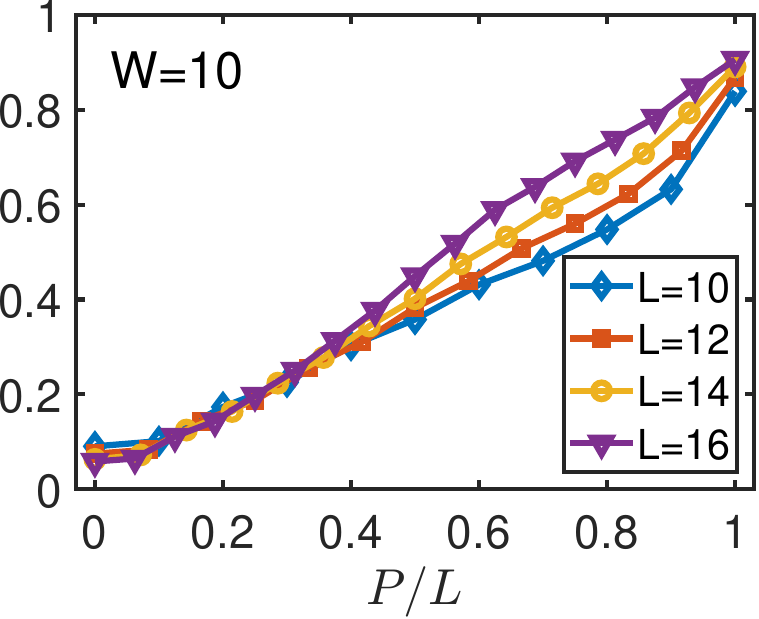} 
\end{centering}
\caption{Normalized EE as a function of $P/L$ for various values of the disorder strength $W$. From the intersection of two curves corresponding to a pair of $L$ and $L+2$ we estimate $(P/L_{av})_c^*$ at a given $W$.}
\label{fig:S6}
\end{figure*}
\begin{figure*}
\begin{centering}
\includegraphics[width=0.235\textwidth]{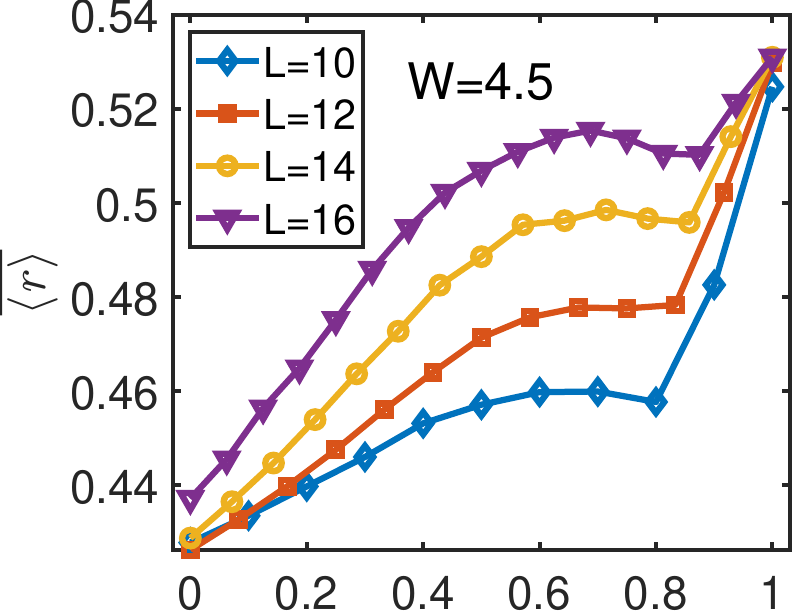} \hspace{0.0mm}
\includegraphics[width=0.225\textwidth]{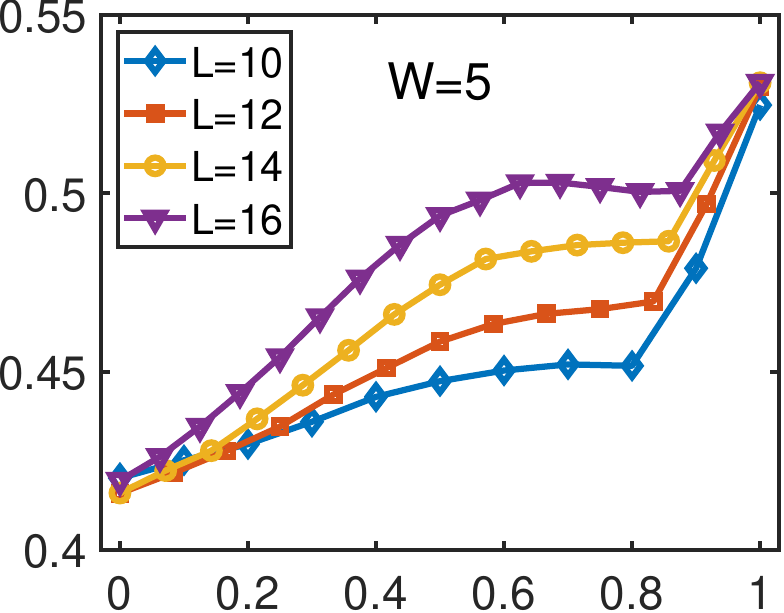}
\hspace{0.0mm}
\includegraphics[width=0.225\textwidth]{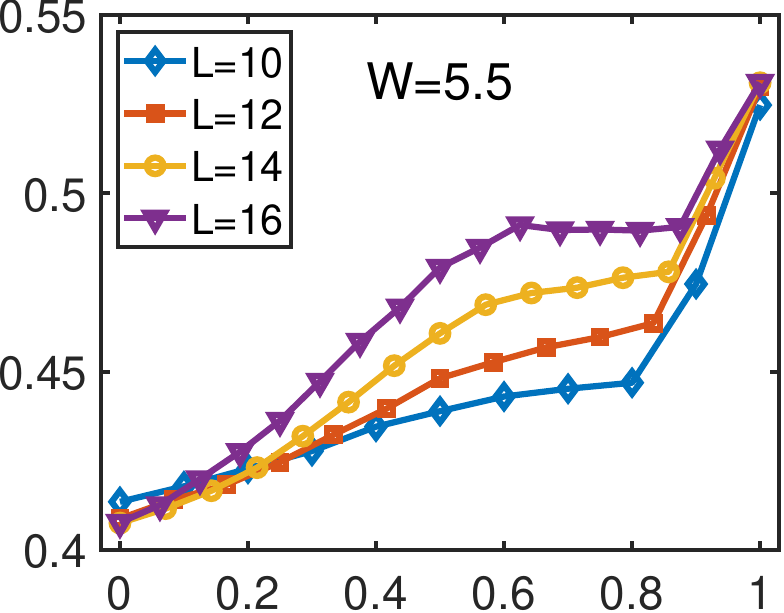}
\hspace{0.0mm}
\includegraphics[width=0.225\textwidth]{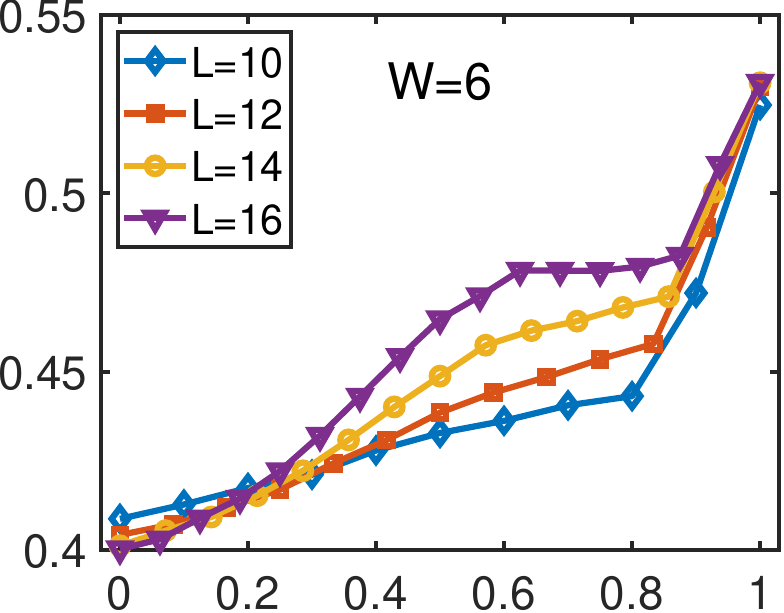} \\
\vspace{2mm}
\includegraphics[width=0.235\textwidth]{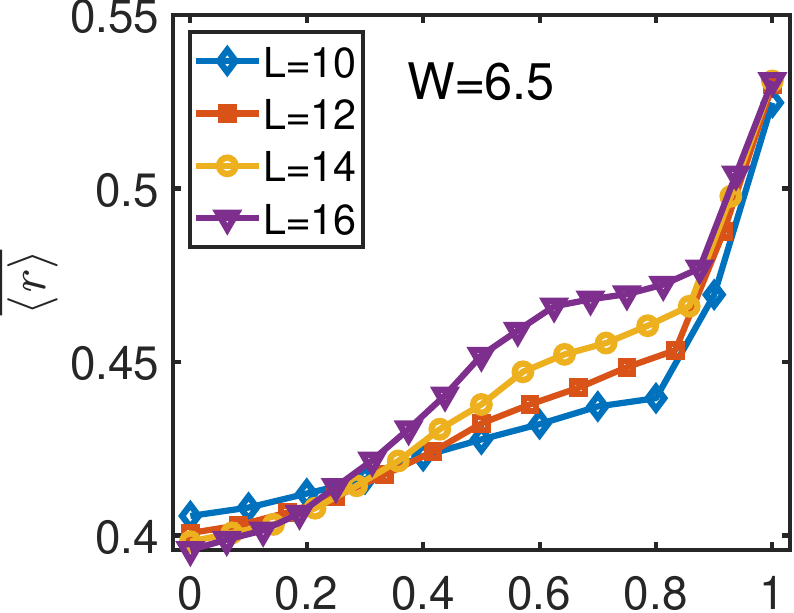} \hspace{0.0mm}
\includegraphics[width=0.225\textwidth]{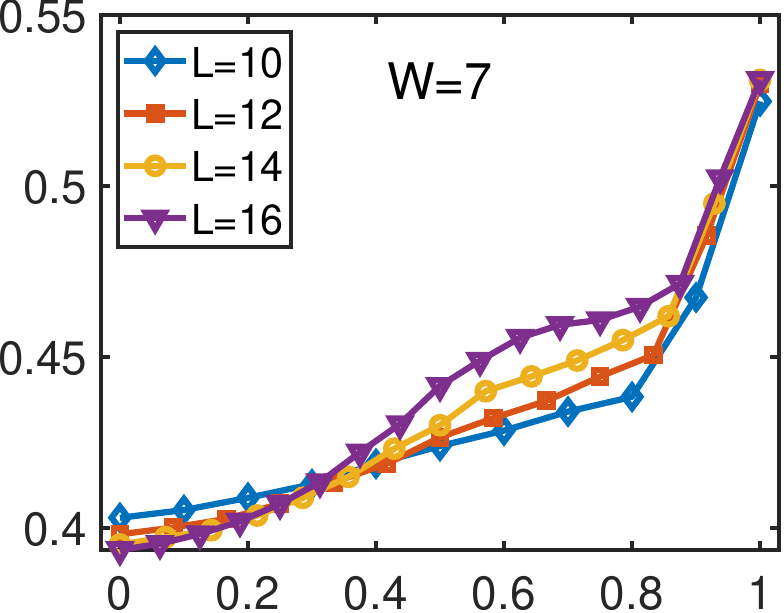}
\hspace{0.0mm}
\includegraphics[width=0.225\textwidth]{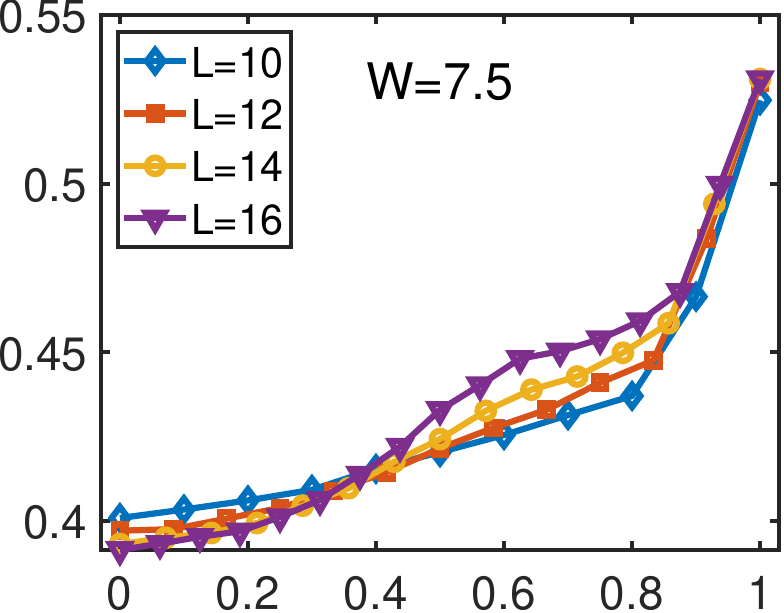}
\hspace{0.0mm}
\includegraphics[width=0.225\textwidth]{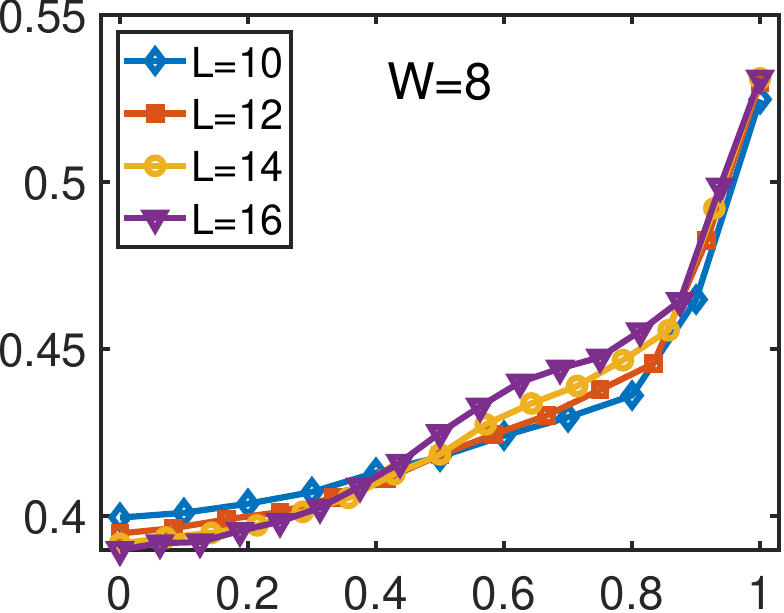} \\
\vspace{2mm}
 \includegraphics[width=0.235\textwidth]{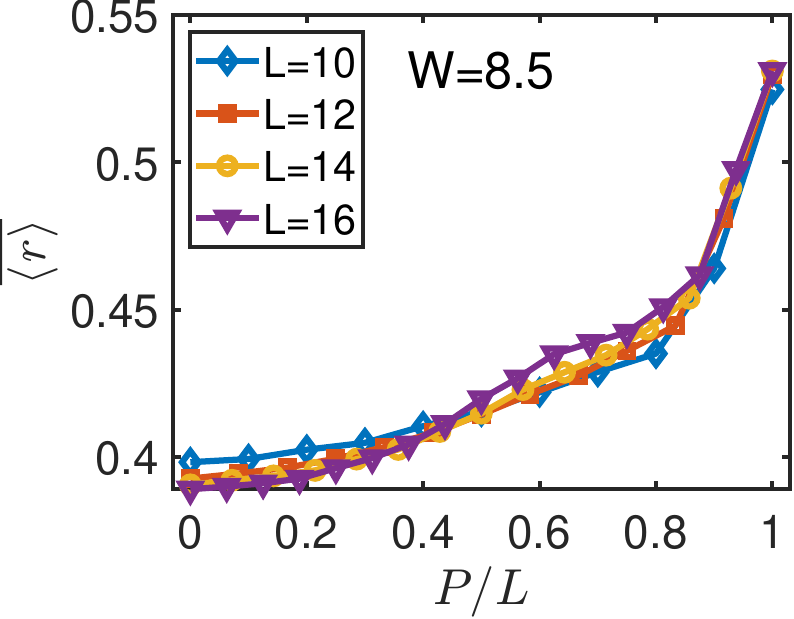} \hspace{0.0mm}
\includegraphics[width=0.225\textwidth]{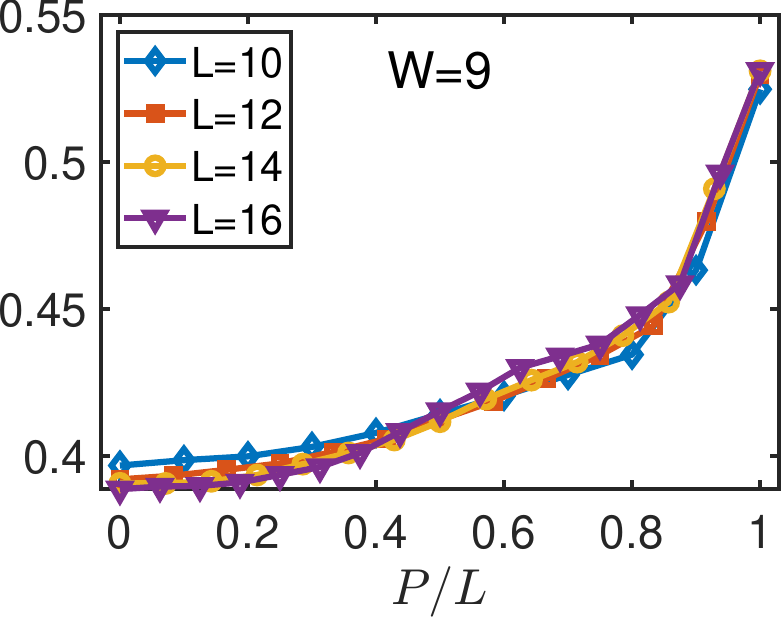}
\hspace{0.0mm}
\includegraphics[width=0.225\textwidth]{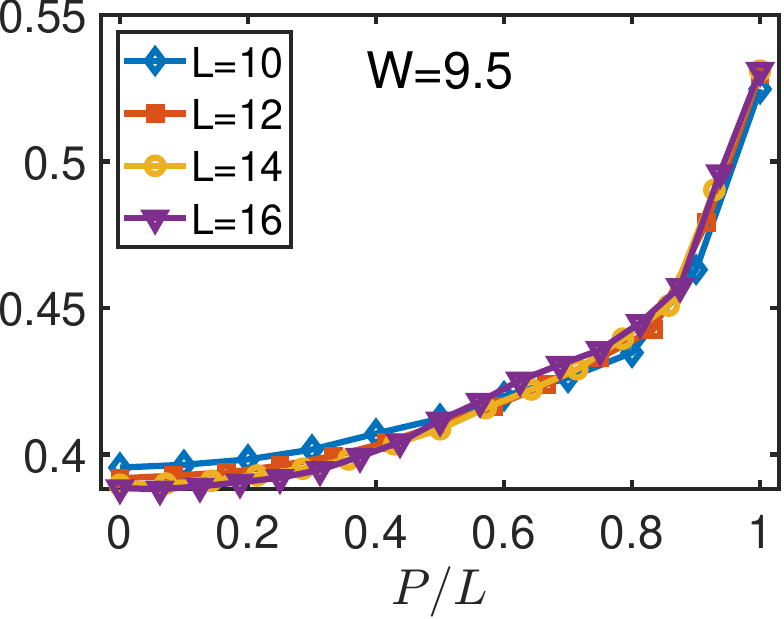}
\hspace{0.0mm}
\includegraphics[width=0.225\textwidth]{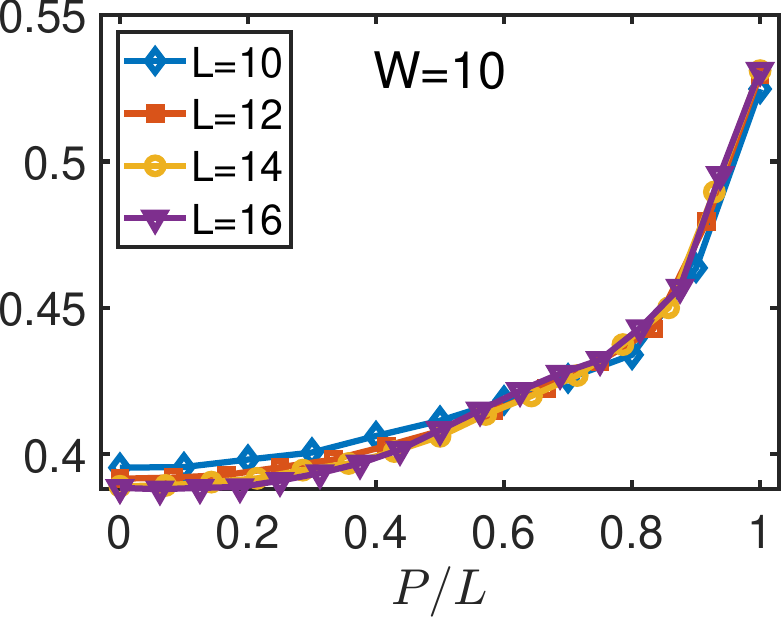} 
\end{centering}
\caption{Averaged GR as a function of $P/L$ for various values of the disorder strength $W$. From the intersection of two curves corresponding to a pair of $L$ and $L+2$ we estimate $(P/L_{av})_c^*$ at a given $W$.}
\label{fig:S7}
\end{figure*}

\section{phase boundary} \label{app.Phase_bounday}
To trace out the phase boundaries $W_c^*(L_{av})$ presented in Fig.~\ref{fig:3} of the main text, for a fixed $P$, we extract the crossing points of the EE or GR when they are plotted as a function of $W$ for curves of size $L$ and $L+2$.
Here $W_c^*(L_{av})$ is the point where the difference between curves of size $L$ and $L+2$ equals zero.
In Fig.~\ref{fig:S4}, we plot $\Delta_S = \overline{ \langle S(L+2) \rangle}/S_{\mathrm{P}}(L+2) - \overline{ \langle S(L) \rangle}/S_{\mathrm{P}}(L)$ as a function of $W$.
In each panel, $P$ is fixed.
For a given $L$ and $L+2$ we identify $W_c^*(L_{av})$ as the point where the corresponding $\Delta_S$ curve crosses the zero line, indicated by the horizontal dashed line.
Similarly, in Fig.~\ref{fig:S5} we have plotted $\Delta_r = \overline{\langle r(L+2)\rangle} - \overline{\langle r(L)\rangle}$ as a function of $W$.
The critical disorder strength $W_c^*(L_{av})$ is estimated using the same method.

To estimate the $(P/L_{av})_c^*$ phase boundaries, we plot a given quantity as a function of $(P/L)$ for a given disorder strength $W$.
The critical point $(P/L_{av})_c^*$ is estimated as the point where a given quantity curve for a system size $L$ intersects the curve for $L+2$.
In the case of EE, $(P/L_{av})_c^*$ corresponds to the points where $\overline{\langle S(L+2) \rangle}/S_{\mathrm{P}}(L+2)$ = $\overline{\langle S(L) \rangle}/S_{\mathrm{P}}(L)$, and for GR these are the points where $\overline{\langle r(L+2) \rangle} = \overline{\langle r(L) \rangle}$.
In Figs.~\ref{fig:S6} and \ref{fig:S7} we have respectively plotted EE and GR as a function of $(P/L)$.
In each panel, $W$ is kept fixed as reported.
For $W{<}W_c^*(L_{av})$, the curves for any given quantity and any pair of ($L$, $L{+}2$) do not cross hence, these are not shown.

The results presented in Fig.~\ref{fig:S6} (Fig.~\ref{fig:S7}) show that for $W{\sim}W_c^*(L_{av})$ on the MBL side, EE (GR) does not exhibit a plateau and increases linearly with $(P{/}L)$. This indicates that localization is not stable against any small low-disorder region for these values of $W$. A plateau begins to develop as $W$ is increased deep inside the MBL phase, which indicates the robustness of the MBL to a certain size range of the low-disorder region.

%

\end{document}